\documentclass[prd,preprint,aps,tightenlines,preprintnumbers,amsmath,amssymb, showpacs,superscriptaddress,nofootinbib,longbibliography]{revtex4-1}

\usepackage[T1]{fontenc}
\usepackage[latin9]{inputenc}
\usepackage{float}
\usepackage{bm}
\usepackage{dcolumn}
\usepackage{color}
\usepackage{amsmath}
\usepackage{amssymb}
\usepackage{graphicx} 
\graphicspath{{Pictures/}}

\newcommand{\be}{\begin{equation}}
\newcommand{\ee}{\end{equation}}
\newcommand{\bea}{\begin{eqnarray}}
\newcommand{\eea}{\end{eqnarray}}

\newcommand{\nc}{\newcommand}
\nc{\rnc}{\renewcommand}
\rnc{\d}{\mathrm{d}}
\nc{\D}{\partial}
\nc{\K}{\kappa}
\nc{\bK}{\bar{\K}}
\nc{\bN}{\bar{N}}
\nc{\bq}{\bar{q}}
\nc{\vbq}{\vec{\bar{q}}}
\nc{\g}{\gamma}
\nc{\lrarrow}{\leftrightarrow}
\nc{\rg}{\sqrt{g}}
\rnc{\[}{\begin{equation}}
\rnc{\]}{\end{equation}}
\nc{\nn}{\nonumber}
\rnc{\(}{\left(}
\rnc{\)}{\right)}
\nc{\q}{\vec{q}}
\nc{\x}{\vec{x}}
\rnc{\a}{\hat{a}}
\nc{\ep}{\epsilon}
\nc{\tto}{\rightarrow}
\rnc{\inf}{\infty}
\rnc{\Re}{\mathrm{Re}}
\rnc{\Im}{\mathrm{Im}}
\nc{\z}{\zeta}
\nc{\mA}{\mathcal{A}}
\nc{\mB}{\mathcal{B}}
\nc{\mC}{\mathcal{C}}
\nc{\mD}{\mathcal{D}}
\nc{\mN}{\mathcal{N}}
\rnc{\H}{\mathcal{H}}
\rnc{\L}{\mathcal{L}}
\nc{\<}{\langle}
\rnc{\>}{\rangle}
\nc{\fnl}{f_{NL}}
\nc{\gnl}{g_{NL}}
\nc{\fnleq}{f_{NL}^{equil.}}
\nc{\fnlloc}{f_{NL}^{local}}
\nc{\vphi}{\varphi}
\nc{\Lie}{\pounds}
\nc{\half}{\frac{1}{2}}
\nc{\bOmega}{\bar{\Omega}}
\nc{\bLambda}{\bar{\Lambda}}
\nc{\dN}{\delta N}
\nc{\gYM}{g_{\mathrm{YM}}}
\nc{\geff}{g_{\mathrm{eff}}}
\nc{\tr}{\mathrm{tr}}
\nc{\oa}{\stackrel{\leftrightarrow}}
\nc{\IR}{{\rm IR}}
\nc{\UV}{{\rm UV}}

\begin{document}

\title{Constraining holographic cosmology using Planck data} 

\author{Niayesh Afshordi} \affiliation{Perimeter Institute for Theoretical Physics, 31 Caroline St. N., Waterloo, ON, N2L 2Y5, Canada} 
       \affiliation{Department of Physics and Astronomy, University of Waterloo, Waterloo, ON, N2L 3G1, Canada}
\author{Elizabeth Gould} \affiliation{Perimeter Institute for Theoretical Physics, 31 Caroline St. N., Waterloo, ON, N2L 2Y5, Canada} 
       \affiliation{Department of Physics and Astronomy, University of Waterloo, Waterloo, ON, N2L 3G1, Canada}
\author{Kostas Skenderis} \affiliation{STAG Research Centre and Mathematical Sciences, Highfield, University of Southampton, SO17 1BJ Southampton, UK}  
       %\affiliation{Mathematical Sciences, Highfield, University of Southampton, SO17 1BJ Southampton, UK}
 
\date{\today}

\begin{abstract}
Holographic cosmology offers a novel framework for describing the very early Universe in which cosmological predictions are expressed in terms of the observables of a
three dimensional quantum field theory (QFT).  This framework includes conventional slow-roll inflation, which is described in terms of a strongly coupled QFT,
but it also allows for qualitatively new models for the very early Universe, where the dual QFT may be weakly coupled. The new  models describe a universe which is non-geometric at early times. %Correspondingly, there are  qualitatively different power spectra of cosmological perturbations, depending on the nature of the dual QFT.
While  standard  slow-roll inflation leads to a (near-)power-law primordial power spectrum, perturbative superrenormalizable QFT's yield a new holographic spectral shape. %The holographic expansion, when compared to the WMAP data, had previously been shown to be viable but disfavoured. 
Here, we compare the two predictions against cosmological observations. We use CosmoMC to determine the best fit parameters, and MultiNest for Bayesian Evidence, comparing the likelihoods.  We find that the dual QFT should be non-perturbative at the very low multipoles ($l \lesssim 30$), while for higher multipoles ($l \gtrsim 30$)
 the new holographic model, based on perturbative QFT,  fits the data just as well as the standard 
power-law spectrum assumed in $\Lambda$CDM cosmology. This finding opens the door to applications of non-perturbative QFT techniques, such as lattice simulations, to observational cosmology on gigaparsec scales and beyond.

%The (naive) holographic expansion over the entire dataset 
%is disfavored by data at low $l$'s at $\sim$2 sigma level (or ``strong'' Bayesian evidence for $\Lambda$CDM).  However, we find that despite having a different primordial power 
%spectrum than the standard , the new models can fit the data equally well 

\end{abstract}

\pacs{}
\maketitle

\section{Introduction}

The current observational data in cosmology are fit very well by the six-parameter $\Lambda$CDM model.
This model is an empirical parametrization for cosmology, 
combining four parameters of the transfer function with two of the primordial
power spectrum. For the transfer function, the parameters correspond to the 
matter contents of the universe, the current rate of expansion of the universe,
and the optical depth (which is related to the time of reionization). This part 
is well-understood in the context of the $\Lambda$CDM framework. The other two parameters, $\Delta_0 ^2$ and $n_s$,  are those of the scalar primordial power 
spectrum $\mathcal{P}(q)$, which is taken to have a power-law form:
\begin{equation}
\mathcal{P} \left( q \right)=\Delta_0 ^2\left(\frac{q}{q_{*}}\right)^{n_{s}-1}, \label{power_law}
\end{equation} 
where $q_*$, the pivot scale, is an (arbitrary) reference scale. 

Typically the primordial power spectrum is explained using slow-roll inflation in which 
the early universe undergoes a phase of rapid accelerated expansion. This is used to explain 
the homogeneity and isotropy of the universe by having the expansion increase the
size of the regions which were in causal contact after the big bang to our entire
visible universe as well as making it look flat by being large enough that the 
curvature is not visible. In addition, starting from the quantum adiabatic vacuum, inflationary models typically predict a primordial power spectrum well approximated by the power-law form (\ref{power_law}). 
While inflation is often considered to be the best scenario to explain cosmological observations, it suffers from shortcomings
such as predictivity and falsifiability, sparking a search for alternative possibilities (e.g., \cite{Brandenberger}).

%As an alternative method for examining the early universe, it has been previously 
%proposed to examine the holographic dual to this phase. %\cite{McFadden:2009fg}. 
One of the main issues one is faced with when calculating the predictions of models for the early universe is that 
quantum gravity effects become relevant, while we do not yet have a full theory of quantum gravity. 
Inflation bypasses
this by requiring that the gravitational coupling is weak enough so that only a quantum field 
theory on curved space-time is required. This may be sufficient for explaining (\ref{power_law}) 
but it still leaves open the question of what happens at earlier times -- inflation does not 
resolve the issue of the initial singularity. Moreover, the theory is still  generically sensitive to UV issues, as radiative corrections can significantly alter the inflationary action.
For these reasons, it is important to embed inflation into a UV complete theory.

%There is another potential way to bypass our lack of knowledge of quantum gravity. 
Insight from the study of black hole
entropy has long indicated that gravity might have a holographic nature \cite{'tHooft:1993gx, Susskind:1994vu}, {\it i.e.}
that there is a dual quantum field theory (QFT) in one lower dimension without gravity. This principle, the 
holographic principle, should also apply to the early universe. Explicit examples of 
holographic dualities were found in string theory \cite{Maldacena:1997re}. However, these cases tend to apply to theories with a negative cosmological constant, which is in contrast to cosmological observations. 

The extension of the duality to de Sitter spacetime and cosmology was considered soon after the initial work on Anti-de Sitter space \cite{Hull:1998vg, Witten:2001kn, Strominger:2001pn, Strominger:2001gp, Maldacena:2002vr}. In the cosmological  context, the statement of the duality is that the partition function of the dual QFT  computes the wavefunction of the universe \cite{Maldacena:2002vr}, using which, cosmological observables may be obtained.  These dualities are less understood than the standard AdS/CFT duality, in part because we currently have no explicit realization in string theory. Nevertheless, one may set-up a holographic dictionary  \cite{McFadden:2009fg, McFadden:2010na, McFadden:2010vh, McFadden:2011kk, Bzowski:2011ab}
using a correspondence \cite{Skenderis:2006jq}  between cosmological accelerating solutions and  holographic renormalization group (RG) flows, solutions that admit standard holographic interpretation.
%On the QFT side, this correspondence amounts to a specific analytic continuation on parameters and momenta. 

In this duality, time evolution is mapped to inverse renormalization group flow and the physics of the early universe is mapped to the IR physics of the dual QFT.
Thus, depending on the nature of the IR, we have different cosmological scenarios. In this paper, we test theories for the very early Universe 
against the cosmic microwave background (CMB) data, so more precisely we would like to know what the dual QFT is which is relevant at the energy scales probed by the CMB. 

One of the main properties of the holographic dualities is that they are strong/weak coupling dualities. This means that when one of the two sides is strongly coupled, and therefore
difficult if not impossible to solve, the other side is weakly coupled, and
solvable perturbatively. Therefore a weakly coupled inflationary period is dual to a
strongly coupled quantum field theory. While work has been done in using  
holography in this setting (see  \cite{Maldacena:2011nz, Hartle:2012qb, Hartle:2012tv,Schalm:2012pi, Bzowski:2012ih, 
Mata:2012bx, Garriga:2013rpa, McFadden:2013ria, Ghosh:2014kba,Garriga:2014ema, Kundu:2014gxa, Garriga:2014fda,
McFadden:2014nta, Arkani-Hamed:2015bza, Kundu:2015xta, Hertog:2015nia,Garriga:2015tea, Garriga:2016poh} 
for a sample of works in this direction)  we here mainly examine the opposite case. This is the case of a strongly coupled gravitational theory. 
In this case, the early universe does not have a well defined geometry. It can not be examined
without quantum gravity. However, the dual QFT not only can be examined, 
but is weakly coupled and solvable perturbatively. This is the alternative model we examine, 
which we call the holographic model or holographic cosmology here.\footnote{As inflation is also holographic, this is a potentially confusing terminology. Here we want to distinguish between a cosmology which has a conventional spacetime description (inflation) and one without such description (holographic cosmology).}
In this case the dual QFT is a super-renormalizable three-dimensional QFT.\footnote{An example of such QFT is the worldvolume theory of coincident D2-branes.  The holography nature of this theory is well established \cite{Itzhaki:1998dd, Boonstra:1998mp, Kanitscheider:2008kd}.} 
This model for the very early Universe was first proposed in \cite{McFadden:2009fg} and it was subsequently analyzed in \cite{McFadden:2010na, McFadden:2010vh, McFadden:2011kk, Bzowski:2011ab,Coriano:2012hd, Kawai:2014vxa}. 
A sketch of the Penrose diagram describing holographic cosmology is shown in Figure 1.

%\begin{figure}
%\includegraphics[width=0.75\paperwidth]{HolographicCosmology}
%
%\caption{\label{fig:hcs}
%Schematic diagram for defining a holographic dual to cosmology.
%}
%\end{figure}

\begin{figure}
\includegraphics[width=0.40\paperwidth]{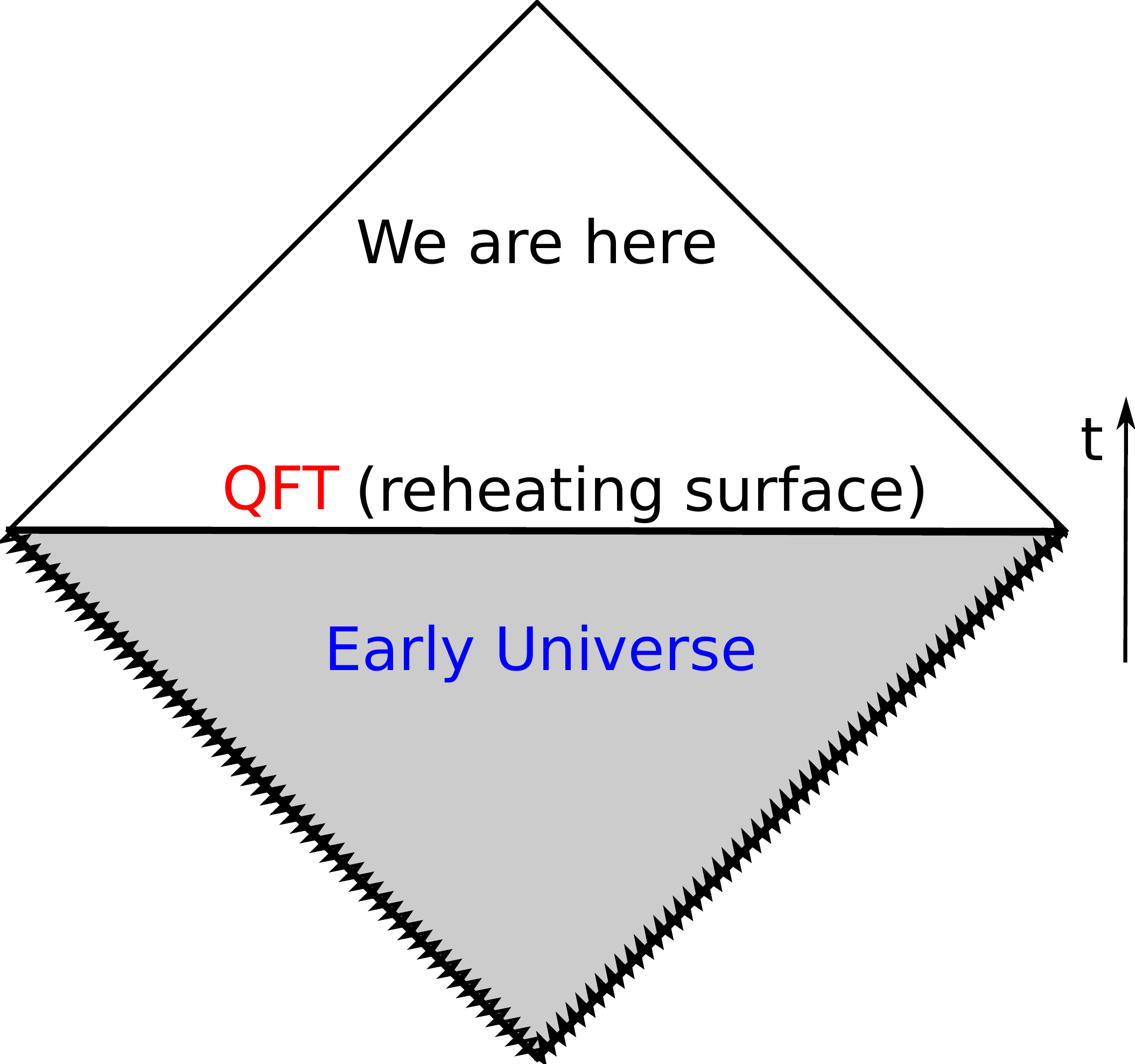}

\caption{\label{fig:pen}
A sketch of the Penrose diagram describing holographic cosmology (HC). The early Universe is non-geometric and is described by a dual QFT, which is located at the end of the non-geometric phase.
}
\end{figure}

Previously, this holographic model had been compared to WMAP7. It was found to be viable \cite{Dias:2011in, Easther:2011wh} but mildly  disfavoured  relative to $\Lambda$CDM.
With the release of the Planck data, it is time to  reexamine the viability of the holographic model for early universe cosmology. Our results were announced in \cite{Afshordi:2016dvb} and the purpose of this paper is to provide a more detailed and comprehensive discussion of their derivation.

The structure of the paper is as follows.  In Section \ref{Models}, we describe the two models we are comparing.
In Section \ref{Data}, we find and explain the  best fit model. Section \ref{Fit} explains how well the two models fit the data
and compare to each other. Finally, in Section \ref{Concl} we present some concluding remarks.

%--------------------------------------------

\section{Models\label{Models}}

\subsection{Holography for cosmology: basics}

The idea of holography for cosmology is that the dual QFT computes the wavefunction of the Universe \cite{Maldacena:2002vr}.
Schematically, this works as follows: The wavefunction is equal to the partition function of the dual QFT,
\begin{equation} \label{wavef}
\psi(\Phi) = Z_{QFT}[\Phi],
\end{equation} 
where $\Phi$  on the left-hand side  denotes collectively gravitational perturbations and on the right-hand side sources that couple to gauge invariant operators.
Note that we consider the wavefunction of perturbations only in this paper.
Cosmological observables may be computed using standard quantum mechanics
\begin{equation} \label{corr}
\langle \Phi(x_1) \cdots \Phi(x_n) \rangle = \int D \Phi |\psi|^2 \Phi(x_1) \cdots \Phi(x_n),
\end{equation}
where the correlators are evaluated at end of the early universe phase (for example, at the end of the inflationary phase, if inflation describes the very early universe).
Using that  $Z_{QFT}[\Phi]$ may be expressed in terms of correlation functions
\begin{equation} \label{zqft}
Z_{QFT}[\Phi]  = \exp \left(\sum_n \frac{(-1)^n}{n!} \langle O(x_1) \cdots O(x_n) \rangle \Phi(x_1) \cdots \Phi(x_n)  \right),
\end{equation}
where $O$ denotes the gauge invariant operators to which $\Phi$ couples.\footnote{We take the QFT to be Euclidean, though this is not essential.} 
We now may express  cosmological observables in terms of QFT correlation functions.
If the QFT is strongly coupled, then the bulk is described by Einstein gravity and these results should match those coming from  standard inflationary cosmological perturbation theory, while if the QFT is weakly coupled the bulk is non-geometric. 

There is currently no first principles derivation of the QFT relevant for cosmology but one may use the domain-wall/cosmology correspondence \cite{Skenderis:2006jq} to map the cosmology problem to that of standard gauge/gravity duality, then use the QFT dual to the domain-wall and finally map the results back to cosmology \cite{McFadden:2009fg,  McFadden:2010na}.\footnote{There is a proposed duality \cite{Anninos:2011ui} where the QFT is defined a priori (i.e. without the need to map the problem to the domain-wall first) but in this case the bulk involves Vasiliev's higher spin gravity instead of Einstein gravity.} This leads to the following holographic formulae for the scalar and tensor spectra, $\mathcal{P}$ and $\mathcal{P}_T$, respectively, 
\begin{equation} \label{holo_2pt}
\mathcal{P}(q) = -\frac{q^3}{16  \pi^2} \frac{1}{{\rm Im} B(q)} , \quad
\mathcal{P}_T(q) = -\frac{ 2 q^3}{\pi^2} \frac{1}{{\rm Im} A(q)},
\end{equation}
where $A, B$ are extracted from the momentum space 2-point function of the energy momentum tensor $T_{ij}$,
\begin{equation} \label{2point}
\langle \langle T_{ij}(q)T_{kl}(-q)\rangle \rangle = A(q)\Pi_{ijkl}+ B(q)\pi_{ij}\pi_{kl} .
\end{equation}
Here $\langle T_{ij}(q_1)T_{kl}(q_2)\rangle = (2 \pi)^3 \delta^3(q_1+q_2) \langle \langle T_{ij}(q_1)T_{kl}(-q_1)\rangle \rangle$, $\pi_{ij} = \delta_{ij} - q_i q_j/q^2$ is a transverse projector and $\Pi_{ijkl}=1/2 (\pi_{ik} \pi_{jl}+\pi_{il} \pi_{jk}-\pi_{ij} \pi_{kl})$ is a transverse-traceless projector. In other words, the scalar power spectrum is associated with the 2-point function of the trace of the energy-momentum tensor while the tensor power spectrum is related with the transverse-traceless part of the 2-point function. These formulae were derived for QFTs that admit a 't Hooft large $N$ limit and they either become conformal in the UV or  approach a QFT with a generalized conformal structure (where generalized conformal structure is explained in the next subsection). The imaginary part in (\ref{holo_2pt}) is taken after the  analytic continuation,
\begin{equation} \label{analytic}
q \to -i q, \quad N \to -i N, 
\end{equation}
where $q$ is the magnitude of the momentum vector and 
we assume that we are dealing with an $SU(N)$ gauge theory coupled to matter in the adjoint representation, as is the case below.\footnote{In the case of a large $N$ vector model, we need $N \to - N$.} 
Similarly, one can relate the  bispectra with 3-point functions of the energy momentum tensor \cite{McFadden:2010vh, McFadden:2011kk, Bzowski:2011ab}.

When the QFT is strongly coupled, the bulk is geometric and there is a conventional description in terms of quasi-de Sitter or power law inflation. In these cases,
(\ref{holo_2pt}) correctly reproduces the results of cosmological perturbation theory  \cite{McFadden:2009fg,  McFadden:2010na}. Here we focus on the opposite regime where the QFT is weakly coupled.

\subsection{Non-geometric models}

In non-geometric models, the theory is defined by giving the dual QFT.  Here we analyze the model proposed in  \cite{McFadden:2009fg,  McFadden:2010na}, in which
the QFT is an $SU(N)$ gauge theory coupled to scalars 
$\Phi^M$ and fermions $\psi^L$, where $M, L$ are flavor indices. 
The action is given by 
%----Define Lagrangian----
\begin{eqnarray} \label{action}
S &=& \frac{1}{g_{\mathrm{YM}}^2} \int d^3 x \, \mathrm{tr} \left[ \frac{1}{2} F_{ij} F^{ij} 
+ \delta_{M_1M_2} \mathcal{D}_i \Phi^{M_1} \mathcal{D}^i \Phi^{M_2} 
+ 2 \delta_{L_1 L_2} \bar{\psi}^{L_1} \gamma^i \mathcal{D}_i \psi^{L_2} \right. \nonumber \\
&& \left.\qquad \qquad \qquad  
+ \sqrt{2} \mu_{M L_1 L_2} \Phi^M \bar{\psi}^{L_1} \psi^{L_2} 
+ \frac{1}{6} \lambda_{M_1 M_2 M_3 M_4} \Phi^{M_1} \Phi^{M_2} \Phi^{M_3} \Phi^{M_4} \right] , %\label{action}
\end{eqnarray}
%-------------------------
where all fields, $\varphi = \varphi^a T^a$,  are in the adjoint of $SU(N)$ and 
$\mathrm{tr} T^a T^b = \frac{1}{2}  \delta^{ab}$. $F_{ij}$ is the Yang-Mills field strength, and 
$\mathcal{D}$ is a gauge covariant derivative. 
The Yukawa couplings $\mu$ and the quartic-scalar couplings $\lambda$  are dimensionless, while $g_{YM}^2$ has dimension 1. 

 This theory is superrenormalizable and has the important property that has a 
``generalized conformal structure.'' This means that if one promotes $g_{YM}^2$ to a new field 
that transforms appropriately under conformal transformation, the theory becomes 
conformally invariant \cite{Jevicki:1998ub, Kanitscheider:2008kd}. Related to this: if one assigns ``4d dimensions'' to the fields, 
$[A]=[\Phi^M]=1, [\Phi^L]=3/2$, then all terms in the action scale the same way. 
%These terms are then "marginal" from the  perspective of the generalized conformal structure. 
While this is not a symmetry 
of the theory, it still has implications.%  (the implications of generalized conformal structure will be further discussed in \cite{CDS}).

In our case, the generalized conformal structure and the large $N$ limit implies that the 2-point function takes the form
\begin{equation}
A(q, N)= q^3 N^2 f_T(\geff^2), \qquad B(q, N)= \frac{1}{4}  q^3 N^2 f(\geff^2)
\end{equation}
where $f_T(\geff^2)$ and $f(\geff^2)$ are (at this stage) general functions of their argument and $\geff^2 = g_{YM}^2 N/q$ is the effective dimensionless 't Hooft coupling constant.
The factor $q^3$ reflects the fact that the energy momentum tensor has dimension 3 in three dimensions and the factor of $N^2$ is due to the fact that we are considering the leading term in the large $N$ limit. The factor of 1/4 in $B$ is conventional. 

Under the analytic continuation (\ref{analytic})
\begin{equation}
q^3 N^2 \to - i q^3 N^2, \qquad \geff^2 \to \geff^2,
\end{equation} 
so for this class of theories one may readily perform the analytic continuation and (\ref{holo_2pt}) becomes
\begin{equation} \label{holo_2pt_gcs}
\mathcal{P}(q) = \frac{q^3}{4 \pi^2 N^2 f(\geff^2)} , \quad
\mathcal{P}_T(q) = \frac{2 q^3}{\pi^2 N^2 f_T(\geff^2)}. 
\end{equation}
We have thus now arrived in a relation between cosmological observables and correlators of standard QFT.

Perturbation theory applies when $\geff^2 \ll 1$. Since $\geff^2= g_{YM}^2 N/q$,  $\geff^2 \to 0$, as $q \to \infty$, reflecting the fact that the theory is super-renormalizable.
On the other hand the effective coupling grows in the IR, so the question of whether the theory is perturbative or not depends on the scales we probe. In the perturbative regime, the functions $f$ and $f_T$ up to 2-loops take the form
\begin{eqnarray} \label{2loop}
f(\geff^2) &=& f_0 \left( 1 - f_1 \, \geff^2 \ln \geff^2 + f_2 \, \geff^2 + O(\geff^4)  \right), \\
f_T(\geff^2) &=& f_{T0} \left( 1 - f_{T1} \, \geff^2 \ln \geff^2 + f_{T2} \, \geff^2 + O(\geff^4)  \right).
\end{eqnarray}
The coefficients $f_0$ and $f_{T0}$ come from 1-loop and have been computed in \cite{McFadden:2009fg,  McFadden:2010na}. The 2-loop computation is discussed in detail in \cite{CDS}.
At 2-loops there are both UV and IR divergences and these induce the log terms. Both $A$ and $B$ suffer from UV divergences and these can be removed with a counterterm. If (some of) the scalars in (\ref{action}) are  non-minimally coupled scalars\footnote{When non-minimal scalars are coupled to gravity, their action contains a coupling to curvature, $\int \xi R \Phi^2$. Correspondingly, their energy-momentum tensor contains a new term proportional to the so-called improvement term.} then the $B$ form factor (but not the $A$) also has an IR divergence.  It is believed that this class of theories is non-perturbatively IR finite, with the Yang-Mills coupling effectively playing the role of an IR cut-off \cite{Jackiw:1980kv, Appelquist:1981vg}. In summary, $f_1$ and $f_{1 T}$ can be computed unambiguously in perturbation theory, while $f_2, f_{2 T}$ are scheme dependent
and $f_{2}$ also has an IR ambiguity.  As discussed in \cite{Afshordi:2016dvb}, we fix the scheme dependence by setting the RG scale $\mu$ equal to the pivot scale $q_*$,  $\mu = q_*$, and the IR ambiguity of $f_2$ by setting the IR cut-off equal to $g_{YM}$.

Following \cite{Easther:2011wh}, we define new dimensionless variables $g, \beta, g_T, \beta_T$ via 
\footnote{This parametrization assumes that $f_1 \neq 0$, $f_{1T} \neq 0$. While generically this is true, there are also examples where this does 
not hold. For example, (\ref{action}) with only scalars has $f_1=0$. These cases require a separate analysis.} 
\be \label{f_g}
f_1 \gYM^2 N = g q_*, \quad \ln \beta = -\frac{f_2}{f_1} - \ln |f_1|, \qquad f_{1T} \gYM^2 N = g_t q_*, \quad \ln \beta_t = -\frac{f_{T2}}{f_{T1}} - \ln |f_{T1}|
\ee
In terms of new variables 
\begin{eqnarray}
\mathcal{P}\left( q \right) =\frac{\Delta_0 ^2}{1+\left( gq_{*}/q\right) \ln \left| q/\beta gq_{*}\right|}% +{\cal O}(gq_{*}/q)^2}, 
\label{eq:hcps} %\\
\qquad \mathcal{P}_T\left( q \right) =\frac{ \Delta^2_{0T}}{1+\left( g_tq_{*}/q\right) \ln \left| q/\beta_t g_tq_{*}\right|}% +{\cal O}(g_Tq_{*}/q)^2}
\end{eqnarray}
where
\be \label{Delta0}
\Delta_0 ^2= \frac{1}{4 \pi^2 N^2 f_0}, \qquad \Delta^2_{0T} = \frac{2}{\pi^2 N^2 f_{T0} }.
\ee
We emphasize that these formulae were derived using perturbation theory, so our first task when fitting to data is to assess whether the perturbative expansion is justified at all scales seen by Planck.
We use as an indication of the breakdown of perturbation theory the size of $gq_{*}/q$. % (but one should keep in mind that a better measure is the size of  $f_1 \, \geff^2 \ln \geff^2$).
Note that, unlike  \cite{Easther:2011wh}, we did not set $\beta=1$. The theoretical computation \cite{CDS} shows that generically $\beta \neq 1$, and furthermore, $\beta=1$
provides a bad fit to the data (see  Figure \ref{fig:gbeta} or Table \ref{tab:bffull}). We are thus
forced to use 3 parameters to fit the primordial spectrum, one more than needed for $\Lambda$CDM in (\ref{power_law}).

Note that the form of the power spectrum (\ref{eq:hcps}) is a universal prediction for this class of theories, so if this form is disfavoured by the data then it rules out this class of holographic models. On the other hand, if  (\ref{eq:hcps}) is consistent with data, one can further analyze whether the best fit values can be reproduced by a specific choice of QFT within this class.

\subsection{Empirical models} \label{sec:empirical}

To formalize the comparison we now define (following  \cite{Easther:2011wh}) the empirical model of holographic cosmology (HC) to be the model parametrized by the seven parameters
 $(\Omega_{b}h^{2}, \Omega_{c}h^{2}, \theta, \tau, \Delta_0^2, g, \ln \beta)$, where
$\Omega_{b}h^{2}$ and $\Omega_{c}h^{2}$ are the  baryon and dark matter densities, $\theta$ is the angular size of the sound horizon at recombination and $\tau$ is the optical depth due to re-ionization.

This model is to be compared with $\Lambda$CDM, which is parametrized by six parameters, $(\Omega_{b}h^{2}, \Omega_{c}h^{2}, \theta, \tau, \Delta_0^2, n_s)$
and $\Delta_0^2$, $n_s$ are the parameters entering in (\ref{power_law}).

We also compare HC with $\Lambda$CDM with running, which includes as a new parameter the running $\alpha_s = d n_s/d \ln q$. In this case the scalar power spectrum is given by
\begin{equation}\label{eq:lcdmps}
\mathcal{P}\left( q \right) =\Delta_0^2 \left(\frac{q}{q_{*}}\right)^{\left(n_{s}-1\right)+\frac{\alpha_s}{2}\ln\left(\frac{q}{q_{*}}\right)}.
\end{equation}
The running is usually set to zero since it does not improve the fitting significantly. Here we include this model so that we can also compare HC to a model with the same number of parameters.

In inflationary models, $n_s$ typically has weak dependence on $q$ and it may be Taylor expanded around $q_*$. In $\Lambda$CDM, one keeps the leading order term in this expansion, while in $\Lambda$CDM with running one keeps in addition the sub-leading term. In slow-roll inflation, the running is second order in slow-roll parameters and higher order running is further suppressed \cite{Kosowsky:1995aa}. The holographic power spectrum (\ref{eq:hcps}) can be rewritten in the form (\ref{power_law}) with specific $n_s=n_s(q)$  when $gq_{*}/q \ll 1$. In this case, however, $\alpha_s/(n_s-1)=-1$, and higher order runnings are not suppressed \cite{McFadden:2010na, Easther:2011wh}. 

All the cosmological parameters other than those quantifying the primordial spectrum --
i.e. those in the transfer function - are the same in all three models. In
addition, all three models have a parameter $\Delta_0^2$ which determines the overall
amplitude of the power spectrum. These parameters are accounted for in
the data analysis using CosmoMC by fitting for $100\theta$,
$\tau$, $\ln\left(10^{10}\Delta_0^2\right)$, $\Omega_{b}h^{2}$, and $\Omega_{c}h^{2}$. 
In addition, all the nuisance parameters of Planck are identical for both
models. The values and details of these are considered irrelevant for the
analysis. For the parameters not shared by the models, $\Lambda$CDM uses $n_s$ and 
$\alpha_s$ if running is included. Holographic cosmology uses $g$ and 
$\ln\left(\beta\right)$. The priors used for the relevant parameters are in Table \ref{tab:priorcmc}.

\begin{table}
\caption{\label{tab:priorcmc} 
Priors for CosmoMC. The priors are the default for CosmoMC for the $\Lambda$CDM parameters.
$g$ and $\beta$ ranges were chosen to ensure viability of the primordial power spectrum. 
}

\noindent\centering{}
\begin{tabular}{|c||c|c|}
\hline 
Parameter & Minimum & Maximum\tabularnewline
\hline 
\hline 
$\Omega_{b}h^{2}$ & $0.005$ & $0.1$\tabularnewline
\hline 
$\Omega_{c}h^{2}$ & $0.001$ & $0.99$\tabularnewline
\hline 
$100\theta$ & $0.5$ & $10$\tabularnewline
\hline 
$\tau$ & $0.01$ & $0.8$\tabularnewline
\hline 
$\ln\left(10^{10} \Delta_0^2\right)$ & $2$ & $4$\tabularnewline
\hline 
\hline
$n_{s}$ ($\Lambda CDM$) & $0.8$ & $1.2$\tabularnewline
\hline 
$\alpha_s$ ($\Lambda CDM$ running) & $-0.05$ & $0.05$\tabularnewline
\hline 
\hline 
$g$ (HC) & $-0.025$ & $-0.001$\tabularnewline
\hline 
$\ln\beta$ (HC) & $-0.9$ & $4$\tabularnewline
\hline 
\end{tabular}
\end{table}

%$\Omega_{b}h^{2}$ and $\Omega_{c}h^{2}$ are the  baryon and dark matter densities, $\theta$ is the angular size of the sound 
%horizon at recombination, $\tau$ is the the optical depth due to re-ionization

%--------------------------------------------

\section{Matching the Model to Data \label{Data}}

\subsection{Best Fit Parameters}

In order to determine how well the models fit to data, we started by finding the best fit parameters,
median and expected ranges using CosmoMC 
\cite{Seljak:1996is,Zaldarriaga:1997va,Lewis:1999bs,Lewis:2002ah,Howlett:2012mh,Lewis:2013hha,camb_notes}.
Because we needed to compare  models with no variations besides the primordial power spectrum,
we ran not only holographic cosmology (for which we needed to modify the code to use our primordial power
spectrum), but also $\Lambda$CDM using the same dataset. We ran $\Lambda$CDM both with and without 
running. Running was used to ensure the likelihoods were compared between models with the same number of 
parameters, while fitting to $\Lambda$CDM without running was done since running has previously been found 
to not make a significant difference \cite{Ade:2015xua}.

We fit the models to two different sets of datasets. For both cases, the datasets used were identical
for holographic cosmology and both $\Lambda$CDM models. The first case is marked as the standard, 
full Planck run, or is not indicated as special. The data sets used in this case were Planck 2015 (low TEB+high $l$ [HM] TT) as well as
lensing \cite{Ade:2015xua,Ade:2015zua,Aghanim:2015wva,Ade:2015fva,Bennett:2012zja,Reichardt:2011yv,Das:2013zf}, 
as well as Baryonic Acoustic Oscillations (BAO) 
\cite{Beutler:2011hx,Blake:2011en,Anderson:2012sa,Beutler:2012px,Padmanabhan:2012hf,Anderson:2013zyy,Samushia:2013yga,Ross:2014qpa} 
and BICEP2-Keck-Planck (BKP) polarization \cite{Ade:2015tva}. The second case, called the high-$l$ run or the 
run without low ls, uses all the same data except does not use the portion of the Planck dataset corresponding
to $l < 30$.

\begin{table}
\caption{\label{tab:bffull}
Planck 2015 and BAO best fit parameters and $68\%$ ranges for
holographic cosmology and $\Lambda$CDM. Data for $\Lambda$CDM
is from a separate run of CosmoMC, included to compare the $\chi^{2}$
numbers. 
}

\noindent \centering{}%
\begin{tabular}{|c||c|c||c|c||c|c|}
\hline
         & \multicolumn{2}{c||}{HC}  & \multicolumn{2}{c||}{$\Lambda$CDM} & \multicolumn{2}{c|}{$\Lambda$CDM with running}\tabularnewline
\cline{2-7} 
                  & best fit & $68\%$ range       & best fit  & $68\%$ range       & best fit  & $68\%$ range      \tabularnewline
\hline 
\hline 
$\Omega_{b}h^{2}$ & $0.02217$&$0.02215\pm0.00021$ & $0.02227$ & $0.02225\pm0.00020$& $0.02231$ &$0.02229\pm0.00022$\tabularnewline
\hline 
$\Omega_{c}h^{2}$ & $0.1173$ &$0.1172\pm0.0012$   & $0.1185$  & $0.1186\pm0.0012$  & $0.1184$  &$0.1186\pm0.0012$  \tabularnewline
\hline 
$100\theta$       & $1.04112$&$1.04115\pm0.00042$ & $1.04103$ & $1.04104\pm0.00042$& $1.04108$ &$1.04105\pm0.00041$\tabularnewline
\hline  
$\tau$            & $0.081$  &$0.082\pm0.013$     & $0.067$   & $0.067\pm0.013$    & $0.069$   &$0.068\pm0.013$    \tabularnewline
\hline 
$10^9\Delta_0^2$  & $2.126$  &$2.126\pm0.058$     & $2.143$   & $2.143\pm0.052$    & $2.151$   &$2.149\pm0.054$    \tabularnewline
\hline 
$n_{s}$           &          &                    & $0.9682$  & $0.9677\pm0.0045$  & $0.9682$  &$0.9671\pm0.0045$  \tabularnewline
\hline 
$\alpha_{s}$      &          &                    &           &                    & $-0.0027$&$-0.0030\pm0.0074$ \tabularnewline
\hline 
$g$               &$-0.0070$ &$-0.0074_{-0.0013}^{+0.0014}$&  &                    &           &                   \tabularnewline
\hline 
$\ln\beta$        & $0.88$   &$0.87_{-0.24}^{+0.19}$&         &                    &           &                   \tabularnewline
\hline 
\hline
$\chi^{2}$        &$11324.5$ &                    & $11319.9$ &                    & $11319.6$ &                   \tabularnewline
\hline  
\end{tabular}
\end{table}

After running CosmoMC to get the distribution of parameters, we ran the minimizer \cite{minimizer} 
included with the code to find the best fit parameters as well as its likelihood. 

This procedure leads to the parameter ranges in Table \ref{tab:bffull} for the best fit and $68\%$ region
of both models using the full Planck dataset. As can be seen, the difference in $\chi^{2}$ is $4.81$. This
means the difference between the models is $2.2\sigma$, favouring $\Lambda$CDM. The difference in likelihood
between $\Lambda$CDM with and without running is less than $1$, so the case with fewer parameters should be
favoured. Our fit for $\Lambda$CDM is comparable to those found by the Planck team.

As mentioned earlier,  the perturbative expansion (\ref{eq:hcps}) requires $\left| (g q^{*})/q \right| \ll 1$.
How large of values of $\left| (g q^{*})/q \right|$ one is willing to accept depends on the error one is willing to tolerate.
%In addition, as mentioned earlier, one should really track $f_1 \g_{eff}^2 \ln \geff^2$. 
Certainly values of $\left| (g q^{*})/q \right|$  which are of order 1 are outside the regime of validity of perturbation theory.  
In our case, as can be seen in Table \ref{tab:bffull}, the best fit value is $g = -0.00703$ and one can check that 
$2 \times 10^{-3} \leq \left| (g q^{*})/q \right| \leq 2.5$, for the multipoles $2500 \leq l \leq 2$ seen by Planck.
Therefore, $\left| (g q^{*})/q \right|$ is indeed very small for almost all multipoles, but it does become large at very low multipoles
(at $l=30$ it is equal to 0.15, at $l=20$ it is 0.25 and at $l=2$ it  is 2.5). It follows that perturbation theory is valid at all scales seen in Planck, except at 
very low multipoles. This is our first major conclusion: the data {\it a posteriori} justify the perturbative treatment for all multipoles but the very low ones.

At very low multipoles one cannot trust the model: a non-perturbative computation of the 2-point function of the energy-momentum tensor is needed in order to work out the predictions of this model for these multipoles.  In order to stay within the regime of validity of the model, we therefore removed the low $l$ data from our dataset and recalculated the parameters. The exact boundary at $l=30$ was determined by the datasets we had from Planck, which offers the data already split between the $l < 30$ and 
$l \ge 30$ data and it is roughly in accordance with the estimate above. In \cite{Afshordi:2016dvb} we further determined which model within the class of (\ref{action}) reproduces the best fit values and within that model one can make a more precise estimate of the point where the perturbative treatment is not justified and this leads to $ l \sim 35$.

\begin{table}
\caption{\label{tab:bfhl}
Same as Table \ref{tab:bffull}, but with $l < 30$ data removed for both holographic 
cosmology and $\Lambda$CDM.}

\noindent \centering{}
\begin{tabular}{|c||c|c||c|c||c|c|}
\hline
         & \multicolumn{2}{c||}{HC}  & \multicolumn{2}{c||}{$\Lambda$CDM} & \multicolumn{2}{c|}{$\Lambda$CDM with running}\tabularnewline
\cline{2-7} 
                  & best fit & $68\%$ range         & best fit& $68\%$ range           & best fit& $68\%$ range          \tabularnewline
\hline 
\hline 
$\Omega_{b}h^{2}$ &$0.02204$ & $0.02202\pm0.00022$  &$0.02227$& $0.02224\pm0.00020$ &$0.02217$& $0.02212\pm0.00024$\tabularnewline
\hline
$\Omega_{c}h^{2}$ & $0.1187$ & $0.1187\pm0.0014$    & $0.1187$& $0.1188\pm0.0013$   & $0.1186$& $0.1188\pm0.0013$  \tabularnewline
\hline
$100\theta$       &$1.04097$ & $1.04099\pm0.00042$  &$1.04108$& $1.04104\pm0.00043$ &$1.04101$& $1.04100\pm0.00041$\tabularnewline
\hline
$\tau$            & $0.067$  & $0.066\pm0.017$      & $0.0703$& $0.068\pm0.016$     & $0.0695$& $0.067\pm0.016$    \tabularnewline
\hline
$10^9\Delta_0^2$  & $2.044$  & $2.043\pm0.074$      & $2.158$ & $2.151\pm0.064$     & $2.151$ & $2.139\pm0.066$    \tabularnewline
\hline
$n_{s}$           &          &                      & $0.9667$& $0.9660\pm0.0048$   & $0.9682$& $0.9666\pm0.0047$  \tabularnewline
\hline 
$\alpha_s$        &          &                      &         &                     & $0.0083$& $0.0090\pm0.0094$  \tabularnewline
\hline 
$g$               &$-0.0130$ &$-0.0127_{-0.0038}^{+0.0042}$&  &                     &         &                    \tabularnewline
\hline 
$\ln\beta$        & $1.01$   &$0.90_{-0.16}^{+0.32}$&         &                     &         &                    \tabularnewline
\hline
\hline 
%$\chi^{2}$        & \multicolumn{2}{c||}{$824.0$}      & \multicolumn{2}{c||}{$824.5$}    & \multicolumn{2}{c|}{$823.5$}   \tabularnewline
$\chi^{2}$        &$824.0$ &                        & $824.5$ &                     & $823.5$ &                    \tabularnewline
\hline 
\end{tabular}
\end{table}

\begin{figure}
\includegraphics[width=0.60\paperwidth]{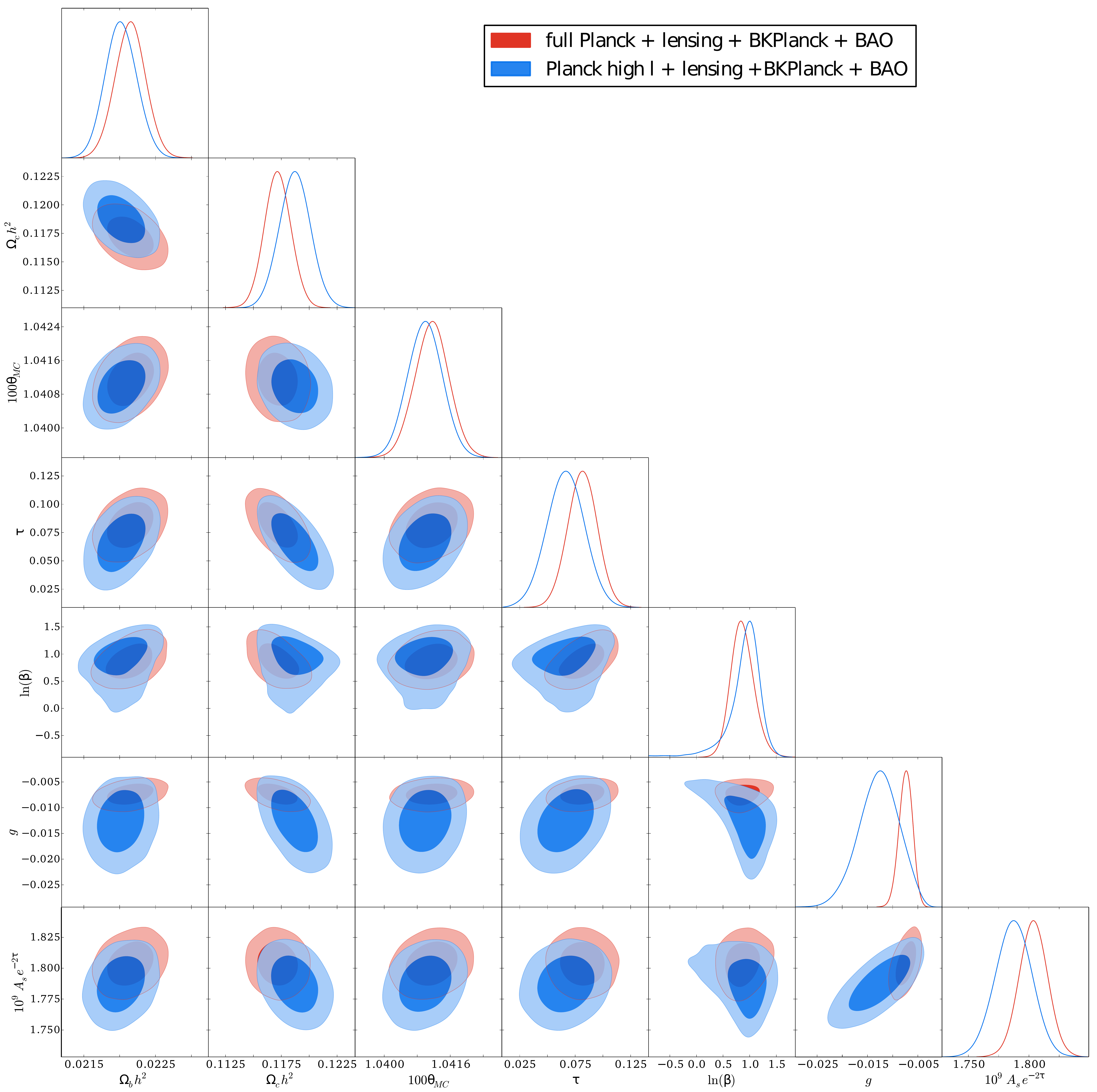}

\caption{\label{fig:tri}A triangle plot of the likelihoods of parameters for holographic cosmology. The blue plots 
showing the case without low $l$s is less symmetric than the red plots with the full data set due to the reduced 
amount of data. The contours show the $68\%$ and $95\%$ confidence levels.}
\end{figure}

Consequently, the results of the new fits can be found in Table \ref{tab:bfhl} if we exclude $l < 30$. For this case, the difference in $\chi^{2}$ is less than $1$,
indicating that the models are within $1.0\sigma$ of each other and that neither model is favoured.
This is our second major conclusion: within their regimes of validity HC and $\Lambda$CDM fit the data equally well.

Figure \ref{fig:tri} shows the shape and degeneracies of the most likely region of parameter space. The most obvious
aspect of these figures is the irregular shape of $\ln\left(\beta\right)$ for the case when the low $l$ data is 
removed. This is seen somewhat in the $1\sigma$ region, but more clearly in the $2\sigma$ region. This seems to
imply that $\ln\left(\beta\right)$ becomes less constrained and potentially consistent with $0$ when the low $l$s are 
removed. The rest of the figure is comparable to Figure 43 of \cite{Aghanim:2015wva}, although the degeneracy between $\Delta_0 ^2$ 
and $g$ is in the opposite direction of that between $\Delta_0 ^2$ and $n_s$ in that figure.

\begin{figure}[b]
\includegraphics[width=0.60\paperwidth]{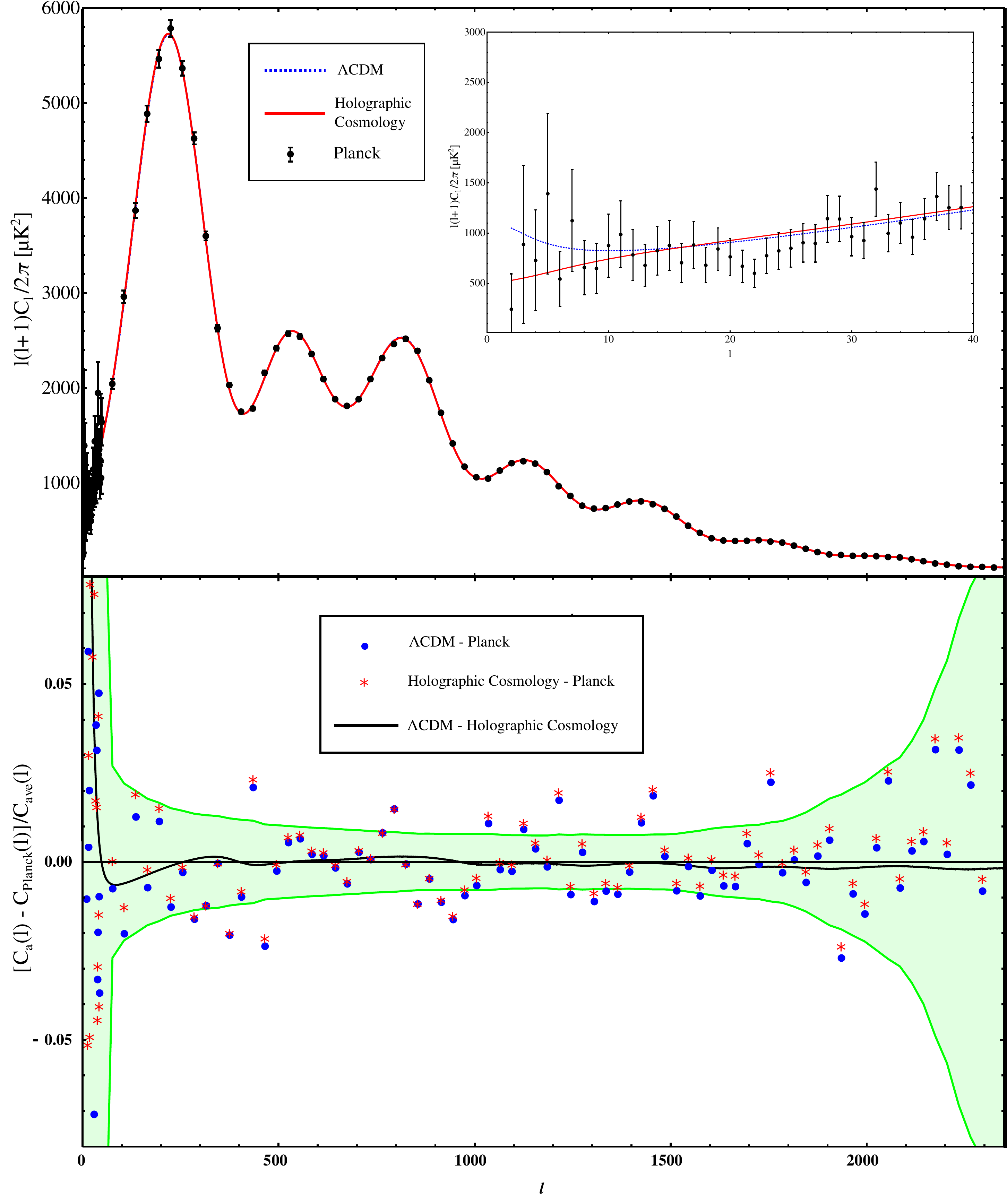}

\caption{\label{fig:tt} TT power spectra of Planck 2015, $\Lambda$CDM and HC. Error bars are shown for low $l$. In the insert ($l\le40$), the blue line ($\Lambda$CDM)
is noticeably above the red one (HC). The green shaded region in the difference 
plot shows the Planck relative error.}
\end{figure}

Taking the parameters from the case with the low-$l$ data removed, we show the TT angular power spectra in Figure 
\ref{fig:tt}  for Planck 2015 data, as well as $\Lambda$CDM and holographic cosmology. Both models
appear to fit the data equally well, with the difference between them being within the $68\%$ region of Planck.
Small $l$'s have the largest difference between the models, however the difference still remains within the error
as low $l$s were not part of calculating the fit.

\begin{figure}

\includegraphics[width=0.375\paperwidth]{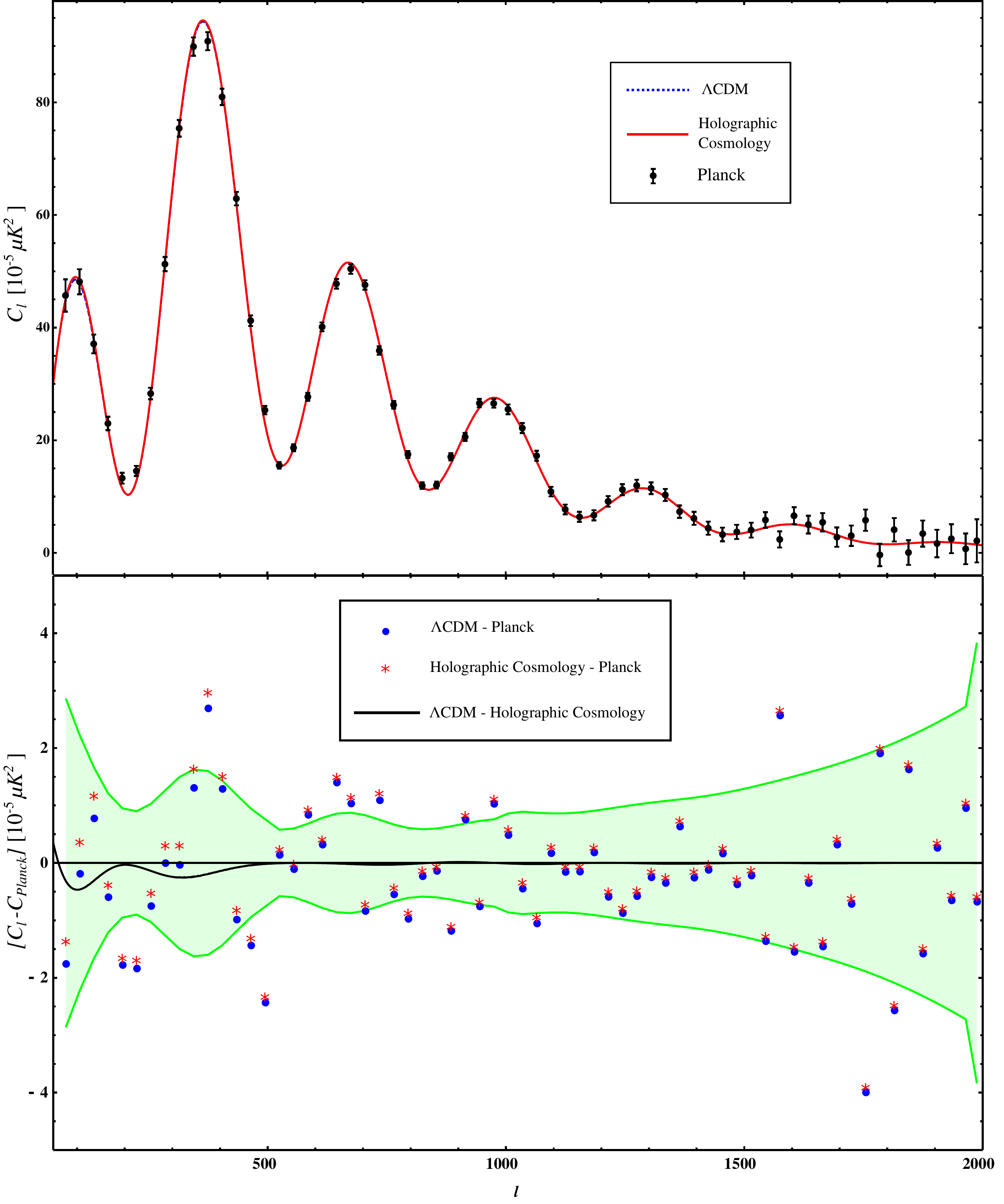}\includegraphics[width=0.38\paperwidth]{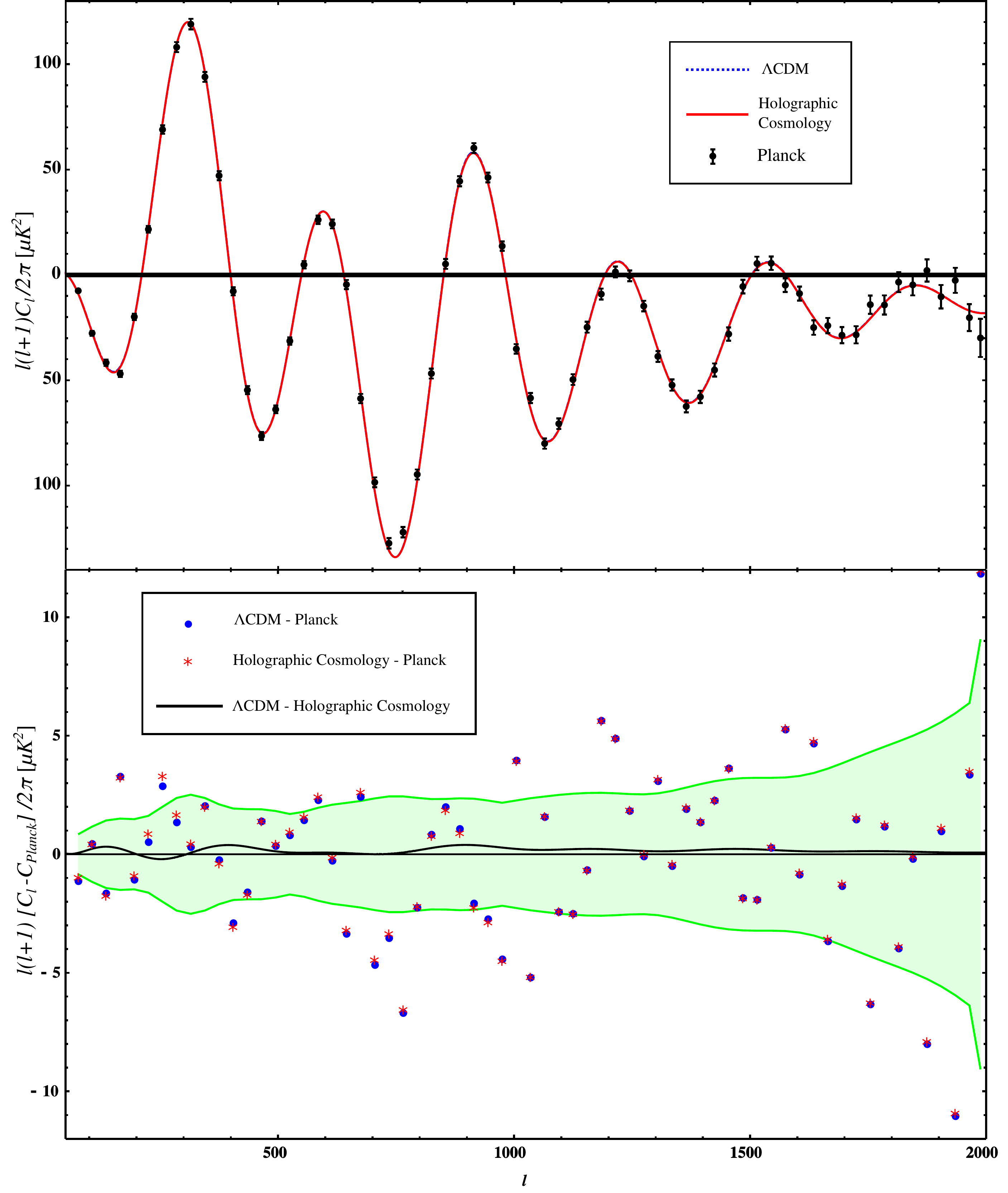}

\caption{\label{fig:ee}Plots of EE (left) and TE (right) polarization for Planck 2015 (black),
$\Lambda$CDM (blue) and HC (red). The green shaded region in the difference 
plot shows the Planck error.}
\end{figure}

Similar plots for the TE and EE power spectra are shown in Figure \ref{fig:ee}.
These plots do not include the low-$l$ data however. The goodness of fit is similar to the TT case. The units for 
the $C_l$'s match those used in \cite{Ade:2015xua}.

%All common parameters of the three models are within $1\sigma$ of each other (with the notable exception of 
%the optical depth $\tau$ \cite{AGS}). 

\subsection{Comparing Primordial Spectra}

\begin{figure}
\includegraphics[width=0.70\paperwidth]{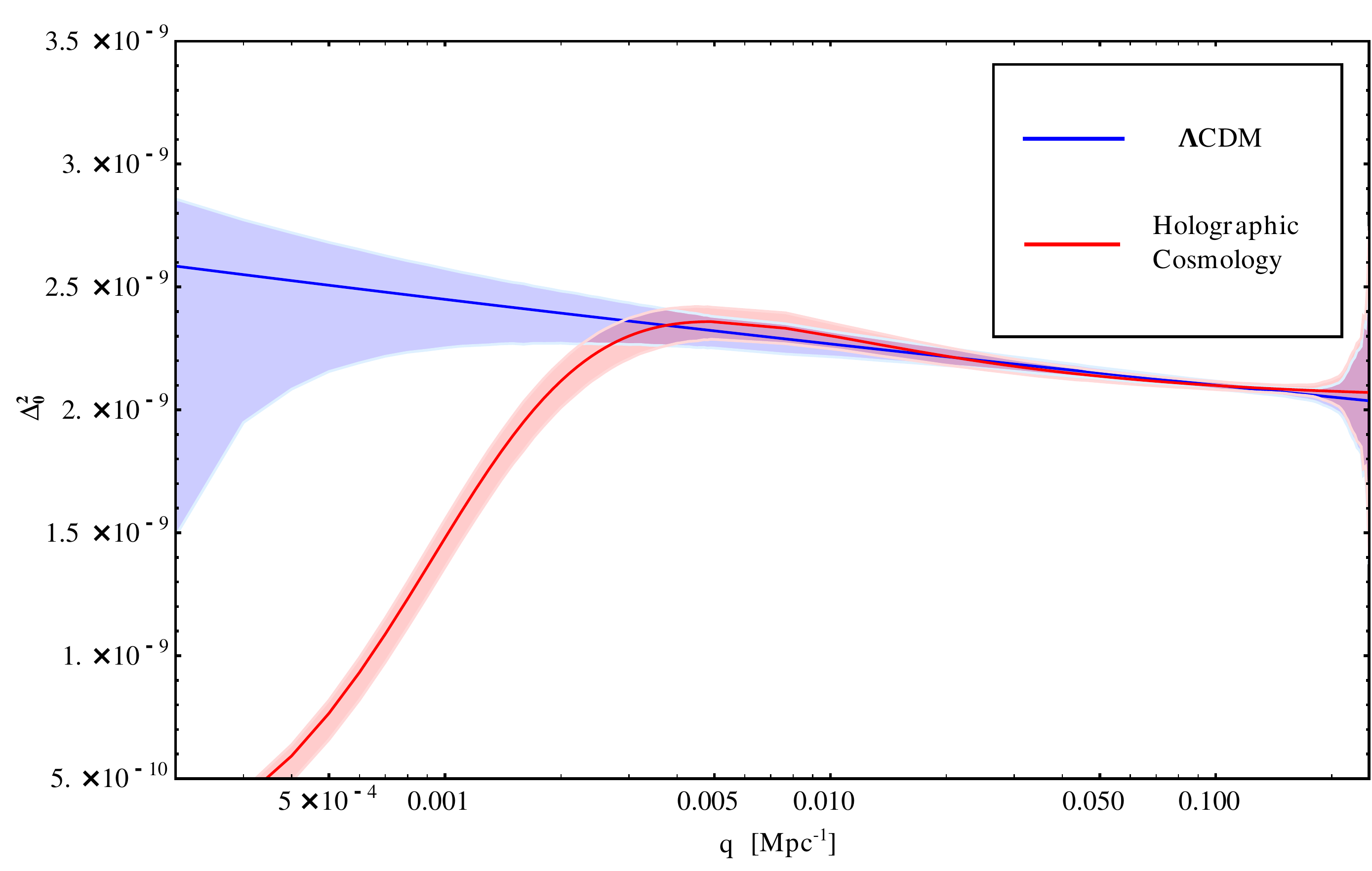}

\caption{\label{fig:prim}
Plot of the primordial power spectrum for HC and
$\Lambda CDM$. The parameters used to produce the curves are the
best fit values in Table \ref{tab:bfhl}. The error (seen in
the lighter shaded regions above and below the curves) is determined
by assuming the same relative error as the Planck cls. It is included
in order to give a sense of the error, not as the actual error. The
red line indicating holographic cosmology starts 
significantly lower and increases rapidly at low $q$ values.}
\end{figure}

Now that we have the best fit parameters, we can examine the difference between the two primordial power spectra. 
This can be seen in Figure \ref{fig:prim}.
We use the best fit parameters for holographic cosmology and $\Lambda$CDM without running found in 
Table \ref{tab:bfhl}. This means we again used the values for when the low $l$ data was removed. The same plot
with the best fit values from Table \ref{tab:bffull} or from much of either tables' indicated range for parameters
would look similar to what is seen. The error is approximated by assuming the same relative error as the Planck TT power spectrum, using $l \approx q \times 14$ Gpc. 

%We can see in Figure \ref{fig:prim} the difference between the two primordial power spectra used. 
The biggest 
difference between the two is seen at low $l$ values. The cutoff used of $l = 30$ is around $q = 0.002 ~{\rm Mpc}^{-1}$. This
removes much of the very low values of the holographic cosmology primordial power spectrum, but still occurs 
(in the middle of the insert) before the holographic cosmology's spectrum has become larger than that of
$\Lambda$CDM. Despite being very similar in value for $q \gtrsim 0.002 ~{\rm Mpc}^{-1}$, the HC and $\Lambda$CDM power spectra can be seen to 
have different shapes.

\begin{figure}
\includegraphics[width=0.70\paperwidth]{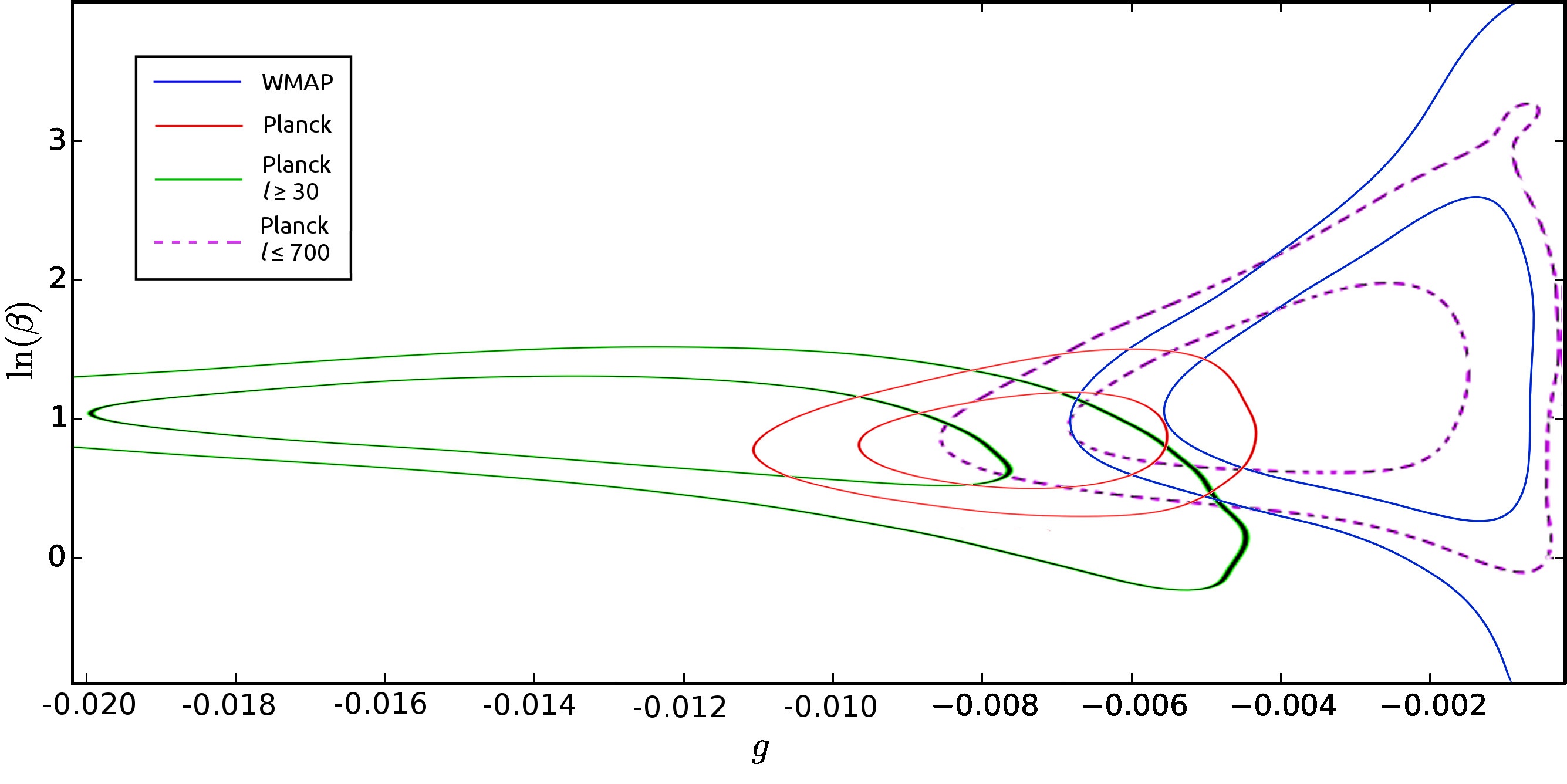}

\caption{\label{fig:gbeta}Plot of $1\sigma$ and $2\sigma$ regions in parameter
space for holographic cosmology $g$ and $\ln(\beta)$ values for
WMAP (blue, right), Planck (red, middle), Planck with low $l$ values
removed (green, left), and Planck with high $l$ values removed (purple,
dashed).}
\end{figure} 

\begin{figure}[b]
\includegraphics[width=0.70\paperwidth]{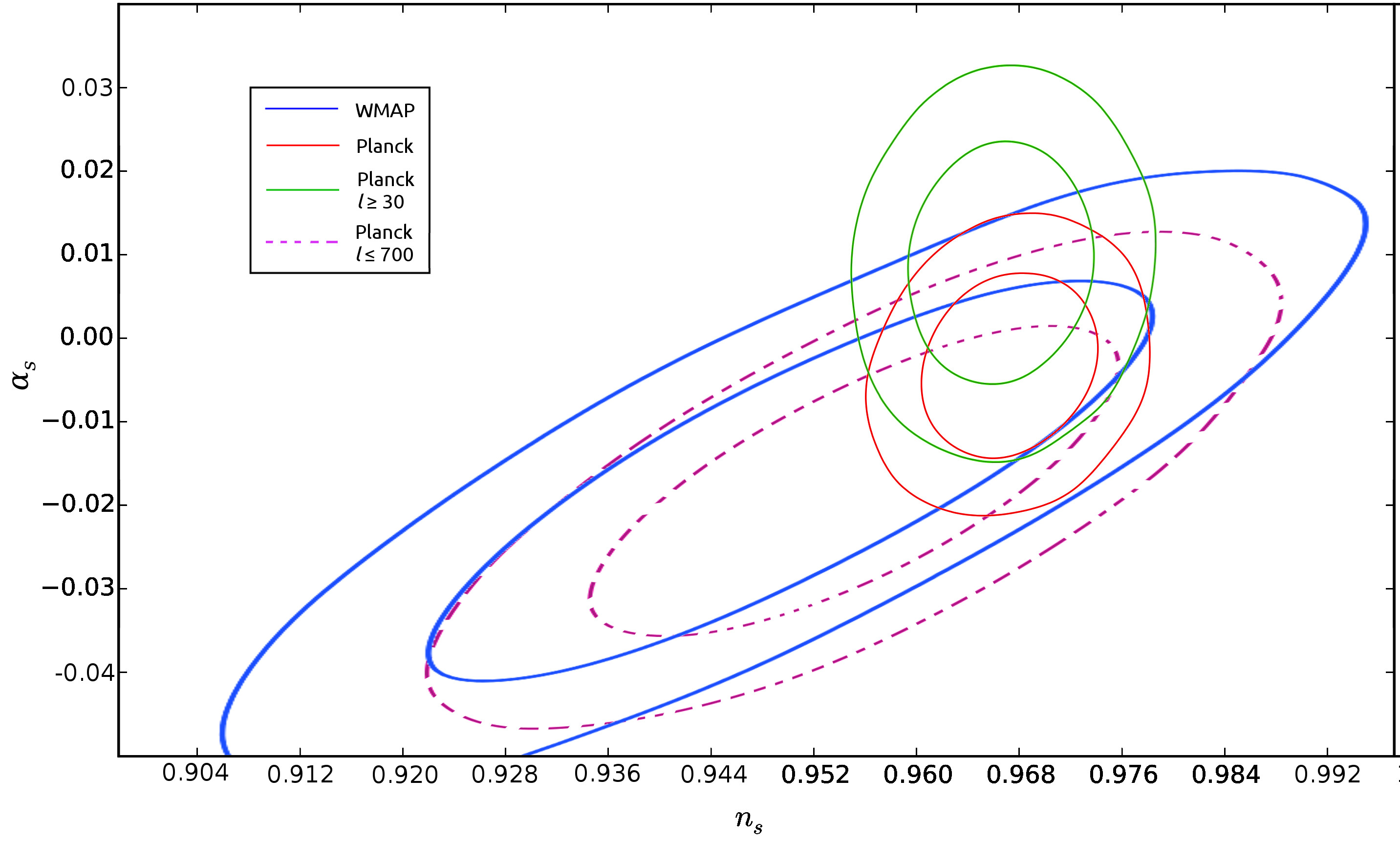}

\caption{\label{fig:nnrun}Plot of $1\sigma$ and $2\sigma$ regions in
parameter space of $\Lambda$CDM $n_s$ and $\alpha_s$ values for WMAP
(blue, largest pair of curves), Planck (red, below the green), Planck
with low $l$ values removed (green, above the red), and Planck with
high $l$ values removed (purple, dashed).}
\end{figure}

\begin{figure}
\includegraphics[width=0.70\paperwidth]{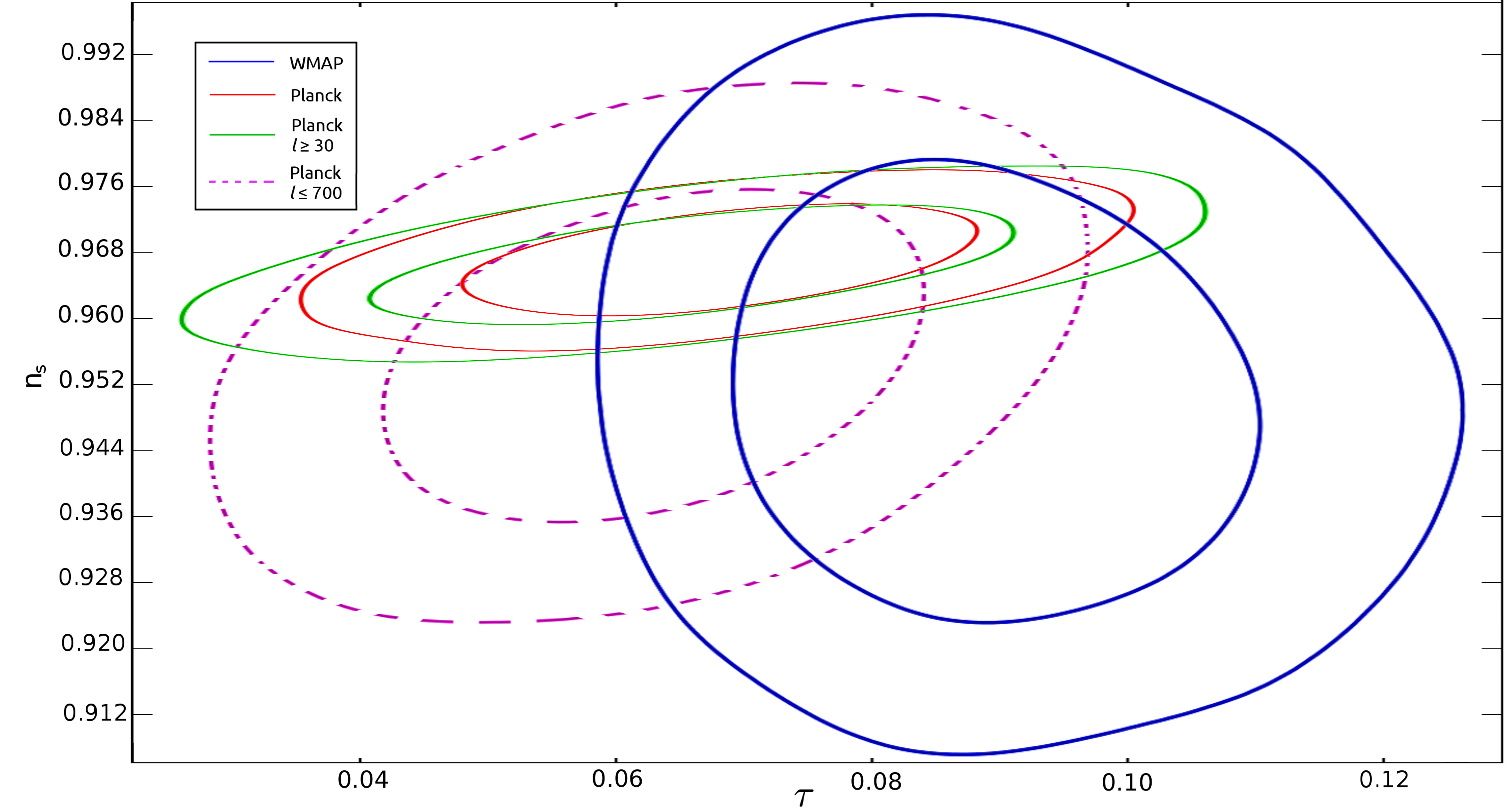}

\caption{\label{fig:ntau}Plot of $1\sigma$ and $2\sigma$ regions in
the parameter space of $\Lambda$CDM $n_s$ and $\tau$ values for WMAP
(blue, largest pair of circles), Planck (red, the smallest set of
circles), Planck with low $l$ data removed (green, slightly larger
than the red), and Planck with high $l$ data removed (purple, dashed).}
\end{figure}

\subsection{Comparison to Previous Results}

%Relative to the WMAP fit in \cite{Easther:2011wh} the value of $g$ has decreased from $-1.3\times 10^{-3}$  to $-7 \times 10^{-3}$. In Fig. \ref{fig:g-vs-ln}, we investigate how the value of 
%$g$ changes if we change the range of multipoles that we consider. It is clear from the plot that the value of $g$ is compatible between WMAP and Planck, if we keep the same multipoles.
%It is also clear that the high $l$ modes want to push $g$ to lower negative values. Larger values of $|g|$ indicate that the theory may become non-perturbative at very low $l$  
%(equivalently, very low $q$) and, as such, the predictions of the model cannot be trusted in that regime. We shall see below that this is supported by model selection criteria.

Comparing the results for WMAP in \cite{Easther:2011wh} to the results for Planck here, it appears that $g$ 
noticeably shifts to lower values, outside of the expected error. This shift remained when %the parameter $\beta$ is added. 
we reran the code for WMAP, this time including $\beta$ and the same external datasets as we used for Planck.\footnote{The parameter $\beta$ was not used in \cite{Easther:2011wh} since it was (incorrectly) argued to be unimportant
for the expected values of $g$. When we ran WMAP again using $\beta$ (not setting it to 1), we got 
$\beta=3.56$ and $g=-0.0027$. These values are used in Figure \ref{fig:gbeta}.} The trend towards more negative values of $g$ 
continues when the low $l$ dataset is removed from the data used to determine the parameters. This trend 
can be seen in Figure \ref{fig:gbeta}. While it is possible that this indicates an issue with the model,
the theory, as stated previously, becomes non-perturbative when $\left|gq_{*}/q\right|$
becomes relatively large. This shift is believed to be compensating for the fact that the model is 
non-perturbative when using the full dataset. To test if the choice of range of $l$s is the reason for the shift in
$g$, we also ran the Planck data without using any data for $l$ above a chosen cutoff of $l=700$ to mimic
the uncertainty in the WMAP data for $l$s around that number. 
\footnote{
Because the data for $l > 30$ is binned every $30$ $l$s, the cutoff point is not exactly $l=700$.
The code is then told to ignore the data for $l$s above the cutoff. The data still remains 
available to be used, however. This makes the cutoff imprecise.
%The information for these higher $l$ values still exists for these runs, but the code is 
%told to ignore this data. This fact, combined with the binning of the data which means $l=700$
%does not  actually exist in the dataset so either the bin below or the bin above must provide the cutoff, 
%makes the cutoff imprecise. 
It is, however, sharper than WMAP, which has data for larger $l$s, but with a very 
large error. See \cite{Planck2016i} for discussions on this type of cutoff.
%Due to complications in how the code works, the cutoff of $l=700$ indicated in the figure was not absolute. 
} Despite the differences in the sharpness of the cutoff, the values found are close to those from WMAP.

A similar analysis for $\Lambda$CDM is shown in Figure 
\ref{fig:nnrun}. For this case, there is no similar shift in $n_s$ and $\alpha_s$.  However, there is a known shift in $\tau$ from WMAP to Planck for the $\Lambda$CDM case: its best fit value went from  0.088 (WMAP) to 0.067 (Planck).
Holographic cosmology with the full dataset gives $\tau =0.081$ which goes down to 0.067 when we remove the $ l <30$ multipoles. The plot of $n_s$ vs 
$\tau$ for $\Lambda$CDM is in Figure \ref{fig:ntau}.

What we can see in these figures is that the shift in $\tau$ for $\Lambda$CDM appears due to Planck while 
the shift in $g$ is at least partially due to the value of $l$. Since $\tau$ decreases to values similar
to Planck when the low $l$ data is removed (Table \ref{tab:bfhl}), it appears that $\tau$ is decreased by Planck, 
but increased to fit the erroneous holographic cosmology power spectrum to compensate for the drop in the low $l$ 
primordial power spectrum. We suspect that the lower $\tau$ value is real.

All other common parameters between the two models are compatible with each other.

\subsection{Tensors}

As in slow-roll inflation, holographic cosmology allows for the production of tensors. There are also
holographic cosmology models consistent with an absence of tensors. The tensor affects
which holographic models are possible, so an analysis of the status of tensors is required. 

\begin{figure}
\includegraphics[height=0.75\paperwidth]{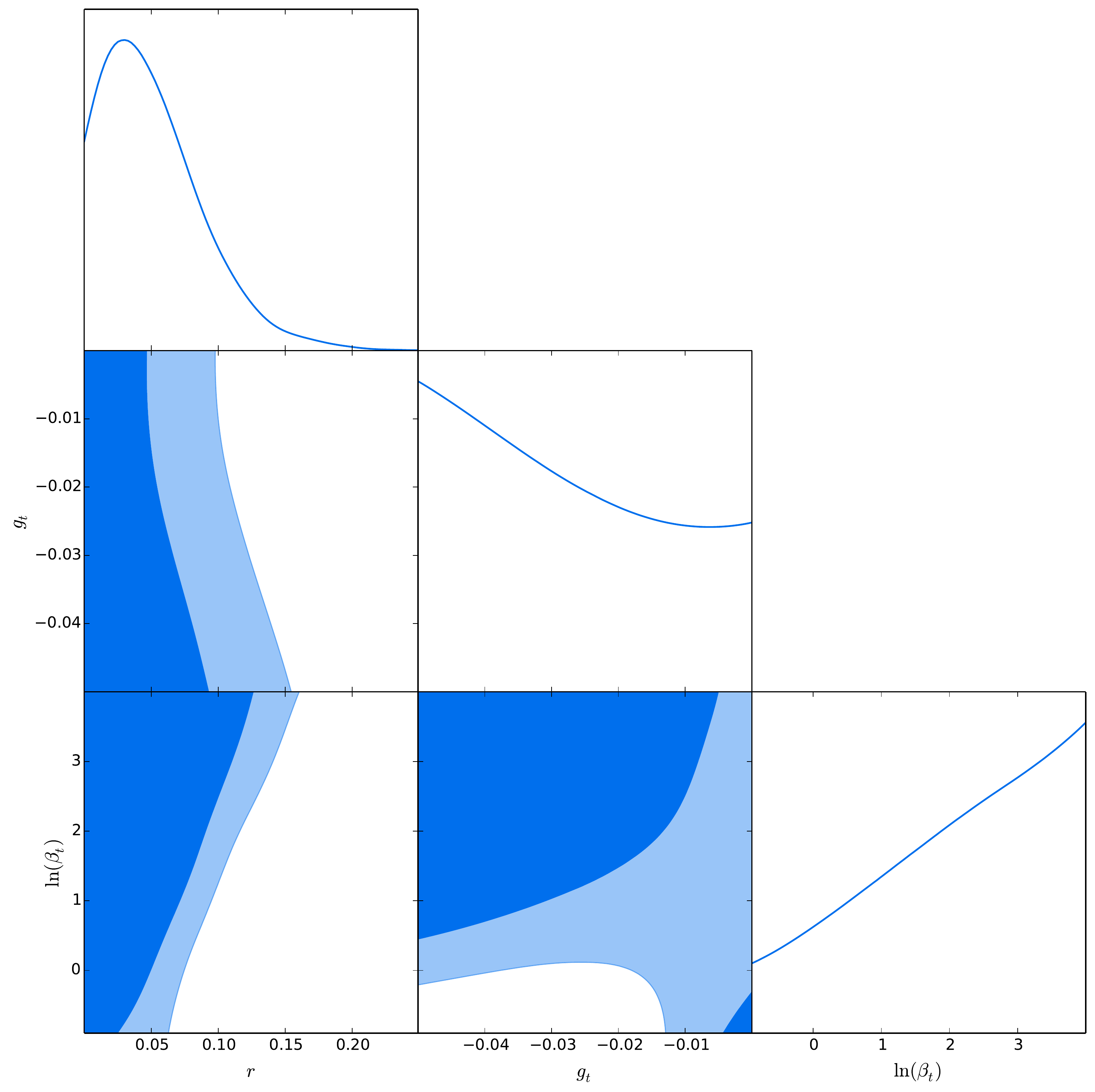}

\caption{\label{fig:tritens}A triangle plot of the likelihoods of parameters for tensors in holographic cosmology. 
The contours show the $68\%$ and $95\%$ confidence levels.}
\end{figure}

In holographic cosmology, the power spectrum for tensors is given in (\ref{eq:hcps}).
%\begin{equation}
%A=rA_{0}\left(\frac{1}{1+\frac{g_{t}q_{*}}{q}\ln\left|\frac{q}{\beta_{t}g_{t}q_{*}}\right|}\right)\label{eq:HCPSa-2}
%\end{equation}
%where $g_{t}$ and $\beta_{t}$ are not necessarily the same as $g$ and $\beta$ for the scalars. 
The upper limit for the ratio of tensors to scalars, $r=\Delta_{0T}^2/\Delta_0^2$, is $12.49\%$ for $2\sigma$ and $17.12\%$ 
for $3\sigma$. The data is consistent with $r=0$. Figure \ref{fig:tritens} shows the triangle plot of
these three parameters, showing that $r=0$ is consistent with the data and consistent with any value of
$g_t$ or $\beta_t$. The allowed value of $r$ can be increased, but this requires the values of 
$\left|g_t\right|$ and $\beta_t$ to be increased past the point for which the perturbative expansion would be
valid. 

%--------------------------------------------

\section{Model Evidence \label{Fit}}

In order to compare different models, one needs to determine which model is more likely given the data. 
Typically one determines which models fit the data better, using for instance the value of $\chi^2$ or 
its square root. While this has already been noted (in Tables \ref{tab:bffull} and \ref{tab:bfhl}), we  examine these likelihoods further here. However, if what we want to know is the probability for the 
model given the data rather than the best fit of the model to the data we should use Bayesian Evidence.
We emphasise that what we compare here are the empirical models introduced in Section \ref{sec:empirical}.

\subsection{Likelihoods}

%Writing this subsection, it seems to be more appropriate for the previous section, to focus on why change
%to using only high ls instead of the chi^2 values.

In order to compare the best fits of the two models, we calculate the difference in $\chi^2$. $\chi^2$ 
is given by $\chi^2 = 2 \left(-\ln \mathcal{L} \right)$, where $\mathcal{L}$ is the likelihood of the model. When we
take the square root of the difference, $\sqrt{\Delta\chi^2}$, we can get the number of standard
deviations one model is from the other. We interpret results within $1\sigma$ as insignificant, but a model is considered to be still viable at up to $3 \sigma$'s. 

However, the likelihood does not account for the number of
parameters in the model. Since we had to include $\beta$
in the holographic cosmology models, we have one more parameter than standard $\Lambda$CDM. 
Instead of adding a term to compensate for the different number of parameters as suggested in 
\cite{Easther:2011wh}, we added running to $\Lambda$CDM so that it has the same number of 
parameters. Increasing the number of parameters decreases the minimum $\chi^2$. Since this decrease, as
seen in Table \ref{tab:bffull}, is less than $1$, the extra parameter is disfavoured in the model.
It does, however, give us a model with the same number of parameters for comparison.

The $\chi^2$ values given in Tables \ref{tab:bffull} and \ref{tab:bfhl} are also presented here in 
Table \ref{tab:like} and \ref{tab:diff} for holographic cosmology and $\Lambda$CDM with running. For the full Planck
dataset, the difference in $\chi^2$ is $4.81$, corresponding to a difference of $2.2\sigma$. 
However, as explained previously, our holographic model breaks down at low $l$ values and cannot be
trusted. Table \ref{tab:like} shows the breakdown in the source of $\chi^2$ based on dataset. As
can be seen, most of this difference comes from low $l$ data, which we do not expect to be accurate.
Comparing instead the model run without the unreliable portions of the data, $\Delta\chi^2=0.5$.
This is within $1\sigma$, indicating that neither model is statistically preferred to the other.

%We thus see that $\Lambda$CDM  and HC do equally well for $l \geq 30$, where 2-loop HC predictions (\ref{eq:hcps}) are reliable.

\begin{table}
\caption{\label{tab:like}$\chi^2$ breakdown for different runs of CosmoMC.
The table shows the $\chi^2$ values of the HC and $\Lambda$CDM with running from 
Table \ref{tab:bffull} (full Planck data) split by dataset. The $\chi^2$'s are split into contributions from the high $l$ dataset ($l \geq 30$), 
the low $l$ dataset ($l<30$) and all other contributions to $\chi^2$. \\
%with the contribution from the low $l$ dataset removed above  the minimum $\chi^2$ values for the $l\ge30$ run from Table \8D for comparison. Given that the best fit for the total data set matches the focused portion of the dataset less well than a best fit for only that portion, there are no obvious issues with the numbers given.
}

\noindent \centering{}%
\begin{tabular}{|c|c|c|}
\hline 
$\chi^2$ breakdown for full Planck run   (Table \ref{tab:bffull}) & HC & $\Lambda$CDM with running\tabularnewline
\hline 
\hline 
Contribution of high $l$ data ($l \geq 30$) & $767.4$ & $766.6$\tabularnewline
\hline 
 Contribution of low $l$ data ($l<30$) & $10498.2$ & $10494.1$\tabularnewline
\hline 
 Contribution of other data & $58.9$ & $58.9$\tabularnewline
\hline 
Total contribution& $11324.5$ & $11319.6$\tabularnewline
\hline 
\end{tabular}
\end{table}

\begin{table}
\caption{\label{tab:diff}$\chi^2$'s, excluding $l < 30$ data, using best-fit parameters from  Tables \ref{tab:bffull} and \ref{tab:bfhl}. }   
\begin{tabular}{|c|c|c|}
\hline
 & HC & $\Lambda$CDM with running\tabularnewline
\hline 
\hline 
$\chi^2$ for full Planck without low $l$ data (from Table \ref{tab:like}) & $826.3$ & $825.5$\tabularnewline
\hline
$\chi^2$ total for $l\ge30$ run (Table \ref{tab:bfhl}) & $824.0$ & $823.5$\tabularnewline
\hline 
\end{tabular}
\end{table}

\subsection{Bayesian Evidence}

In the previous subsection, we added a parameter (running) to $\Lambda$CDM 
in order to have models with the same number of parameters when we use likelihood to 
compare them. In this subsection, we use a method that automatically accounts for the number of parameters:
we compute the Bayesian evidence, the probability of each model given the data (rather than that of the data given the model). 
A detailed exposition of this method can be found in \cite{Jaynes,Trotta:2008qt, Bayesian} and references therein.
As reviewed in \cite{Easther:2011wh}, application of Bayes' theorem leads to 
%\begin{equation}
%P\left(a|b\right)=\frac{P\left(b|a\right)P\left(a\right)}{P\left(b\right)}, \label{eq:bayes}
%\end{equation}
%where $P\left(a|b\right)$ indicates the probability of $a$ given $b$. Here we will choose $a$ to be the
%parameters of the model while $b$ is the data we are attempting to match. In this case, $P\left(b|a\right)$
%is the likelihood which we have already calculated. In this case, however, we will focus on the term in 
%the denominator. This is the probability of the data given the model, calculated by the integral for model evidence:
\begin{equation}
E=\int d\alpha_{M}P\left(\alpha_{M}\right){P\left(D|\alpha_{M}\right)},
\end{equation}
where $\alpha_{M}$ is the set of parameters that specify the model  and $D$ is the data. Here, $P\left(D|\alpha_{M}\right)$ is the probability for obtaining the data $D$ given parameters $\alpha_M$, which is the same as the likelihood $\mathcal{L\left(\alpha_{M}\right)}$ calculated previously. $P\left(\alpha_{M}\right)$ is the prior probability for the parameters.

Our aim is to compare the two empirical models introduced in Section \ref{sec:empirical} and in order to be maximally agnostic 
about the underlying physical models we proceed by using  flat priors, i.e. $P\left(\alpha_{M}\right) =$ const. for all values of $\alpha_M$ which we consider viable, while it vanishes otherwise. %Then the prior probability  is constant over some defined region and zero outside it, and 
Then, the evidence integral becomes
\begin{equation} \label{final_Bayes}
E=\frac{1}{{\rm Vol}_{M}}\int_{{\rm Vol}_{M}} d\alpha_{M}\mathcal{L\left(\alpha_{M}\right)},
\end{equation}
where the integral is over the region of the parameter space in which the prior probability distribution is non-zero and 
${\rm Vol}_{M}$ is the volume of this region.

Alternatively, one could consider comparing physical models, for example a specific inflationary model versus
the model specified by (\ref{action}).  In this case, the prior probabilities would (in principle) be theoretically computable from the underlying model. 
For the case of the holographic model in (\ref{action}) the parameters $g$ and $\beta$ are related by a 2-loop computation 
to the parameters of the underlying model (the rank of the gauge group, the field content, the couplings etc.) and assuming that all perturbative models are {\it a priori} equally  likely\footnote{Alternatively, one may use the partition function of the QFT (with no sources turned on) in order to assign different probabilities to different perturbative models.} 
one can, in principle, compute the prior probability for the parameters $g$ and $\beta$ by analyzing how often given values of $g$ and $\beta$ are realized. 
It would be interesting to see whether such analysis would lead to non-trivial prior distribution.
We leave such analysis to future work and proceed with flat priors, as is common.

%The integral giving the probability of the data given the model can be inverted using Equation 
%\ref{eq:bayes} again. In this case, the probability is easily inverted knowing that $P\left(d\right)$ 
%is the same for both of these models and only the ratio of the probabilities is required. Since the 
%prior assumption is that neither model is favoured, the evidence integral is taken to be the 
%evidence for the model or the Bayesian evidence.
% - using cmc chains to calculate, results
% - using multinest to calculate, standard technique, results, plots, ...

%As discussed in \cite{Easther:2011wh}, if we assume flat priors for all parameters $\alpha_M$ that 
%define a given model
%where $\mathcal{L\left(\alpha_{M}\right)}$ is the likelihood and Vol${}_M$ is the volume of the region 
%in parameter space over which the prior probability distribution is non-zero. 

To compute (\ref{final_Bayes}), we used
MultiNest \cite{Feroz:2007kg,Feroz:2008xx,Feroz:2013hea}. The priors  are determined from the previous fits of the same empirical models to data
%CosmoMC calculation
%Since CosmoMC samples points approximately based on their likelihood, we can make a rough calculation using
%the CosmoMC chains.
%so using the CosmoMC chain, this reduces
%to
%\begin{equation}
%\frac{1}{E}=\frac{1}{N}\sum_{i=1}^{N}\frac{1}{\mathcal{L}_{i}}.\label{eq:evidence}
%\end{equation}
%Here, $E$ is the evidence, $\mathcal{L}$ is the likelihood and $N$
%is the number of elements in the chain. The quantity $-\ln\left(E\right)$
%which is used to compare likelihoods is calculated with this by 
%\begin{equation}
%-\ln\left(E\right)=\ln\left[\frac{1}{\sum_{j=1}^{n}w_{j}}\sum_{i=1}^{n}\frac{w_{i}}{e^{-\left[-\ln\left(like\right)\right]_{i}}}\right].\label{eq:evidence-1}
%\end{equation}
%The calculated numbers are in Table \ref{tab:baycmc}.
% priors
and  are given in Table \ref{tab:priormn}. These priors were chosen to be consistent
with the choices in \cite{Easther:2011wh}. However, the range of $100\theta$ needed to be increased to allow
for the known best fit values. In addition, the range of $g_{min}$ needed to be increased to match the lower 
values of $g$. The range of $g$ was chosen to be $g_{min} < g<0$, with $g_{min}$ variable.
The upper limit was set to $0$ as $g$ was found to be negative in 
\cite{Easther:2011wh} (and the theoretical computation \cite{Afshordi:2016dvb, CDS} also shows that  $g$ is generically negative).  
The maximum $|g_{min}|$  reflects our expectation about the validity of the perturbative expansion. 
We allow for the possibility that the perturbative expansion is valid only for $l>30$. We use as a rough estimate for the validity of perturbation theory that $g q^*/q$ is sufficiently small, taking this to mean a value between 0.20 and 1 at $l=30$.\footnote{The momenta and multipoles are related via $q= l/r_h$, where  $r_h=14.2$ Gpc is the comoving radius of the last scattering surface.} This translates into $-0.009 < g_{\rm min}< -0.45$. The prior for $\beta$ is fixed by using the results from (our fit to) WMAP data. We use two sets of priors: one coming from the 1$\sigma$ range ($0 \leq \ln \beta \leq 2$)  and the other from the 2$\sigma$ range ($-0.2 \leq \ln \beta \leq 3.5$). 
The prior for the running was taken to be  $|\alpha_s| \leq 0.05$. This contains the $1\sigma$ region of $\alpha_s$
for all $1\sigma$ values of $n_s$ for WMAP. It also contains up to the $2\sigma$ region for $\alpha_s$ 
independent of other parameters. Both this and the case with no running were calculated for $\Lambda$CDM.

\begin{table}
\caption{\label{tab:priormn}
Priors used with MultiNest. $g_{min}$
is variable and ranges from $-0.009$ to $-0.65$.
The priors are identical to those used for WMAP \cite{Easther:2011wh}
except for $100\theta$ and $g$ which needed to be expanded to accommodate the 
best fit results and $\beta$ and $\alpha_s$ which were not used originally.
}

\noindent\centering{}
\begin{tabular}{|c||c|c|}
\hline 
Parameter & Minimum & Maximum\tabularnewline
\hline 
\hline 
$\Omega_{b}h^{2}$ & $0.02$ & $0.025$\tabularnewline
\hline 
$\Omega_{c}h^{2}$ & $0.09$ & $1.25$\tabularnewline
\hline 
$100\theta$ & $1.03$ & $1.05$\tabularnewline
\hline 
$\tau$ & $0.02$ & $1.5$\tabularnewline
\hline 
$\ln\left(10^{10}*\Delta_0 ^2\right)$ & $2.9$ & $3.3$\tabularnewline
\hline 
\hline 
$n_{s}$ ($\Lambda$CDM, asymmetric) & $0.92$ & $1.0$\tabularnewline
\hline 
$n_{s}$ ($\Lambda$CDM, symmetric) & $0.9$ & $1.1$\tabularnewline
\hline 
$\alpha_s$ ($\Lambda$CDM running) & $-0.05$ & $0.05$\tabularnewline
\hline 
\hline 
$g$ (HC) & $g_{min}$ & $0$\tabularnewline
\hline 
$\ln\beta$ (HC, small) & $0$ & $2$\tabularnewline
\hline 
$\ln\beta$ (HC, large) & $-0.2$ & $3.5$\tabularnewline
\hline 
\end{tabular}
\end{table}

% results

\begin{figure}[h]
\includegraphics[width=0.75\paperwidth]{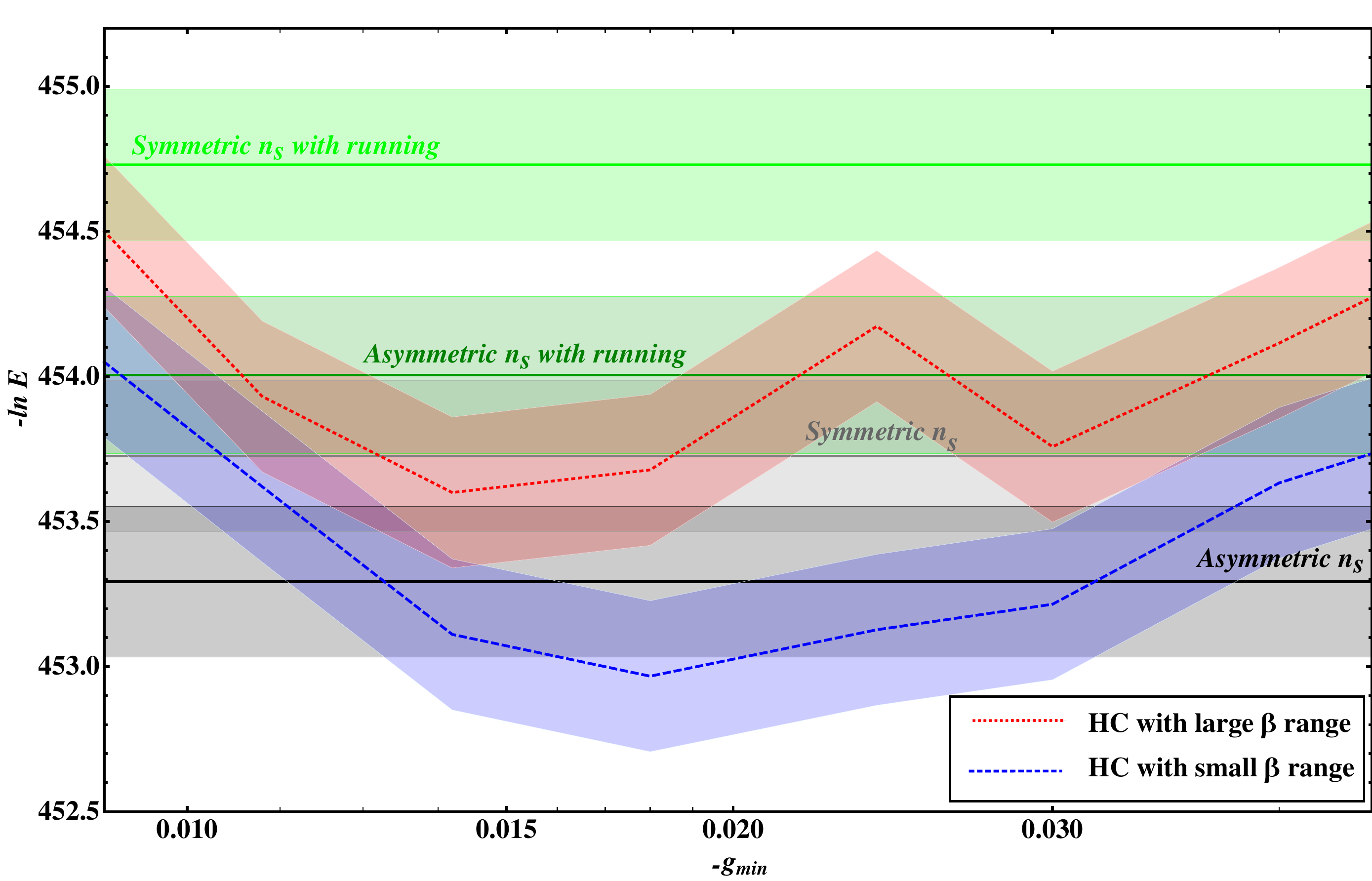}

\caption{\label{fig:bayhl}Plot of Bayesian evidence when $l<30$ data
is removed. Priors are given in Table \ref{tab:priormn}.}
\end{figure}

%The results of the Bayesian evidence MultiNest runs for the full Planck dataset is in Figure \ref{fig:bay}.
In Figure \ref{fig:bayhl}, we present the results for the Bayesian evidence using the data without the low multipole
and for different choices of priors. We use the data without the $l<30$ multipoles because only for this portion of the data the holographic model
is perturbative.  The shading around each line indicates the error.
As a guide \cite{Trotta:2008qt}, a difference $\ln E<1$  is insignificant and $2.5 < \ln E< 5$ is strongly significant. 
We can see that the difference in evidence between $\Lambda$CDM and holographic cosmology is insignificant.

%\begin{figure}
%\includegraphics[width=0.75\paperwidth]{baysian5}
%
%\caption{\label{fig:bay}Plot of Bayesian evidence. Priors are given in
%Table \ref{tab:priormn}.}
%\end{figure}

%--------------------------------------------

\section{Conclusion and outlook \label{Concl}}

In this paper, we confronted holographic cosmology against Planck CMB anisotropy data, as well as other cosmological observations. In this work, holographic cosmology is the empirical model 
obtained by replacing  the primordial power-law power spectrum assumed in $\Lambda$CDM by that obtained (holographically) by a perturbative computation in 
a three-dimensional superrenormalizbale QFT with generalized conformal structure.
We found that the data  {\it a posteriori}  justifies the use of perturbation theory for all but the very low-multipoles ($l<30$).
Restricting to this part of the data, we further found that this theory fits just as well as $\Lambda$CDM.  This follows both from the goodness of fit (the difference of $\chi^2$ is less than 1)
and Bayesian evidence (the difference in log Bayesian evidence is less than one). If  we (incorrectly) use the holographic model over the entire data, then the model is viable but disfavoured. 

In order to include in the analysis the low-multipole data one would need a non-perturbative evaluation of the 2-point function of the energy momentum tensor.
One way to do this is to put the QFT on lattice and use the methods of lattice gauge theory; such computation is currently in progress.
Such non-perturbative results would allow us to meaningfully compare this model with $\Lambda$CDM over the entire data, and may potentially explain the large angle anomalies in the CMB sky (e.g., \cite{Ade:2013nlj}). A lattice computation would also allow us to formulate yet another new class of the holographic models, namely ones based on a QFT with a coupling constant of intermediate strength.  Such models could potentially provide an even better fit than the models we analysed.

In the analysis in this paper we assumed an instant reheating: the data from the end of the very early universe phase were the initial conditions for hot big bang cosmology. It would be useful to develop a dynamical model describing the transition from the non-geometric phase to Einstein gravity. This may be achieved by adding irrelevant operators that would modify the UV sector of the QFT and induce an RG flow that would drive the theory to strong coupling. Such terms could modify the high $l$ part of the spectrum, but our ability to fit the current data very  well 
without such corrections suggests that that they are small.  However, future  results from the next generation stage IV CMB experiments \cite{Abazajian:2016yjj}, as well as future large scale structure surveys such as SPHEREX \cite{Dore:2014cca}, are expected to reach up to much higher wavenumbers,  potentially probing the  holographic reheating phase in our model.

\begin{acknowledgments}
We would like to thank Raphael Flauger for collaboration at early stages of this work.
KS is supported in part by the Science and Technology Facilities Council (Consolidated Grant ``Exploring the Limits of the Standard Model and Beyond'').
NA and EG were supported in part by the University of Waterloo, Natural Sciences and Engineering Research Council of Canada (NSERC), and Perimeter Institute for Theoretical Physics. Research at Perimeter Institute is supported by the 
Government of Canada through the Department of Innovation, Science and Economic Development Canada and by the Province of Ontario through 
the Ministry of Research, Innovation and Science. 
This project has received funding from the European Union's Horizon 2020 research and innovation programme under the Marie Sk\l{}odowska-Curie grant agreement No 690575. 
We acknowledge the use of the Legacy Archive for Microwave Background Data Analysis (LAMBDA), part of the High Energy Astrophysics Science Archive 
Center (HEASARC). HEASARC/LAMBDA is a service of the Astrophysics Science Division at the NASA Goddard Space Flight Center.
\end{acknowledgments}

%--------------------------------------------

%\bibliographystyle{unsrt}
\bibliography{HC}

\end{document}